\documentclass[aps,prr,superscriptaddress,longbibliography,
reprint]{revtex4-1}
\usepackage{graphicx}
\usepackage{bm}
\usepackage{amsmath,amssymb, mathrsfs}
\usepackage{braket,comment}
\usepackage{tikz}
\usepackage{xcolor,units}
\usepackage{amsfonts,dsfont}

\begin{document}

\title{%
	Nonlinear quantum gates for a Bose-Einstein condensate%
}
\author{Shu Xu}
\affiliation{%
	Shanghai Branch, Hefei National Laboratory for Physical Sciences at Microscale, University of Science and Technology of China, Shanghai 201315, China%
}
\email{xushu91@mail.ustc.edu.cn}
\author{J\"org Schmiedmayer}
\affiliation{%
	Vienna Center for Quantum Science and Technology, Atominstitut, TU Wien, Stadionallee 2, 1020 Vienna, Austria%
}
\author{Barry C. Sanders}
\affiliation{%
	Shanghai Branch, Hefei National Laboratory for Physical Sciences at Microscale, University of Science and Technology of China, Shanghai 201315, China%
}
\affiliation{%
    	Institute for Quantum Science and Technology, University of Calgary, Alberta, Canada T2N 1N4%
}
\date{\today}

\begin{abstract}
Quantum interferometry and quantum information processing have been proposed for Bose-Einstein condensates~(BECs),
but BECs are described in complicated ways such as using quantum field theory
or using a nonlinear differential equation.
Nonlinear quantum mechanics does not mesh well with the superposition principle at the heart of interferometry and quantum information processing but could be compatible.
Thus, we develop a rigorous foundation for quantum gates,
obtained by solving the equation for evolution,
and then we employ this foundation,
combined with quantum-control techniques
and appropriate state-sampling techniques,
to devise feasible nonlinear Hadamard gates
and thereby feasible, i.e., high-contrast,
nonlinear Ramsey interferometry.
Our approach to BEC interferometry and quantum logic shifts the paradigm by enlarging to the case of nonlinear quantum mechanics,
which we apply to the cases of BEC interferometry and quantum information processing.
\end{abstract}
\maketitle
\section{Introduction }
\label{sec:intro}
Bose-Einstein condensates (BECs)~\cite{anderson1995} have been proposed as a platform for quantum sensing~\cite{pelegri2018quantum} and for quantum computing~\cite{byrnes2012},
but BECs are rather complicated objects,
described properly in the context of quantum field theory or effectively by the relatively simple but inherently nonlinear Gross-Pitaevskii equation~\cite{pitaevskii1961,gross1961}.
The notion of a quantum gate and the concept of quantum information processing are challenging in the context of either the second-quantized or the nonlinear quantum theories.
Our aim is to formalize quantum gates and quantum information processing in the context of nonlinear quantum mechanics~\cite{pang2005quantum},
apply this theory to controlling and measuring a BEC qubit
(effective two-level system with other energy levels considered negligible),
describe how to define, effect and characterize performance of nonlinear quantum gates,
and to design and then simulate deployment of a nonlinear Hadamard gate to our introduction of a nonlinear type of a Ramsey interferometer~\cite{ramsey1949}.

Specifically, we have the following claims.
First, we define a nonlinear quantum gate as a nonlinear unitary operator that preserves the inner product.
Usually a nonlinear unitary operator preserves the norm but not necessarily the inner product.
Then we use quantum-control methods to determine a feasible shaking trajectory~\cite{bucker2013,van2016optimal}
for a quartic potential confining the BEC,
with the quantum-control target being an evolution that effects a good approximation to a Hadamard gate,
characterized by average gate fidelity over all possible input states.
After solving for feasible potential parameters and for the shaking trajectory,
we apply these methods to a simulation of nonlinear Ramsey interferometry and calculate the resultant contrast.

Our analysis builds on some pillars of knowledge.
The first pillar concerns nonlinear quantum mechanics,
which necessarily relaxes inner-product preservation to norm preservation for unitary operators and of course sacrifices the superposition principle~\cite{schwartz1997}.
The second pillar is the treatment of BEC dynamics using the Gross-Pitaevskii equation~(GPE),
which is a nonlinear equation of motion~\cite{pitaevskii1961,gross1961}.
We also build on an experiment that shakes the BEC,
and realizes a nonlinear Ramsey interferometer,
but with the limitation that its pair of nonlinear Hadamard gates,
punctuated by intervening nonlinear free evolution,
are optimized only for a restricted set of input states~\cite{van2014}.
Furthermore,
our work relies on quantum-control techniques,
whose basic principle is to add external control to the system's free, or ``drift,'' Hamiltonian,
and can be expressed as a feasibility problem~\cite{spiteri2018}.

The structure of our paper is as follows.
We provide salient background in~\S\ref{sec:background}.
Then we describe our approach in~\S\ref{sec:approach}.
We show our results in~\S\ref{sec:results} and discuss these results in~\S\ref{sec:discussion}.
Finally, we present our conclusions in~\S\ref{sec:conc}.
\section{Background }
\label{sec:background}

In this section,
we present salient background for our work on nonlinear quantum gates for BECs.
In~\S\ref{subsec:BECdynamics}, we discuss essentials of BEC dynamics.
Then, in~\S\ref{subsec:nonqm},
we discuss the essential properties of nonlinear quantum mechanics.
Finally, in~\S\ref{subsec:gatefidelity}
we discuss the fidelity of quantum gates and the relevance of fidelity to assessing gate performance.

\subsection{BEC dynamics}
\label{subsec:BECdynamics}

Now we review experimental and theoretical studies about BECs in a one-dimensional trap.
We begin by discussing an effective one-dimensional BEC.
We follow this explanation by presenting the mathematics that describes the evolution of such a one-dimensional BEC.
Finally, we discuss the potential and how it is shaken for controlling the BEC.

We review experimental realizations of one-dimensional BECs.
Experimentally one-dimensional BECs are realized in highly elongated traps, where the confinement in the transverse direction is much tighter then in the longitudinal (axial) direction.
When both the temperature and the interaction are smaller than the transverse confinement scale then motion in the transverse direction freezes out and dynamics in the longitudinal direction dominates.
Nevertheless the effects of the frozen-out dimensions remain~\cite{salasnich2002,gerbier2004quasi}.
One-dimensional BECs have for example been realized in~\cite{kinoshita2004observation,esteve2006observations}.

We now provide a mathematical description of the one-dimensional BEC~\cite{kruger2010weakly,van2014},
which is expressed in the mean-field approximation in terms of the GPE,
\begin{equation}
\label{eq:1dham}
\text{i}\hbar\partial_t\psi(x;t)
=\hat{H}\left(\psi,\lambda;t\right)\psi(x;t)
\end{equation}
for $\partial_t:=\partial/\partial_t$.
In standard quantum mechanics,
the Hamiltonian is a linear operator
on (infinite-dimensional) Hilbert space~$\mathscr{H}$,
which means that it is a homomorphism on a vector space.
The particle wavefunction~$\psi(x;t)$
relates to the Hilbert-space vector~$\ket{\psi(t)}$
according to~$\langle x|\psi(t)\rangle$.
For quantum information processing,
we typically restrict Hilbert space to be of finite dimension~$d$
and denote Hilbert space with restricted dimension by~$\mathscr{H}_d$.

In the nonlinear case presented in Eq.~(\ref{eq:1dham}),
this Hamiltonian is not a linear operator.
For convenience, we simplify derivative operators as
\begin{equation}
\label{eq:partialderivatives}    
\partial_x
    :=\nicefrac{\partial}{\partial_x},\,
\partial_{xx}
    :=\nicefrac{\partial^2}{\partial_x^2}.
\end{equation}
With this notation,
we express the Hamiltonian~(\ref{eq:1dham})
as
\begin{equation}
\label{eq:Hdef}
\hat{H}\left(\psi,\lambda;t\right)
:=-\frac{\hbar^2}{2m}\partial_{xx}+V\left(x-\lambda(t)\right)
+g\left|\psi(x;t)\right|^2,
\end{equation}
with
\begin{equation}
\label{eq:g}
g=N\frac{2\hbar^2a_\text{s}}{ma_\text{r}^2};
\end{equation}
the effective one-dimensional nonlinear coefficient~$N$ is the number of atoms in the BEC,
$a_\text{s}$ characterizes the scattering amplitude and~$a_\text{r}$ 
the oscillator length in the radial direction~\cite{salasnich2002},
and~$V(x)$ is the external time-dependent trapping potential.
Here~$m$ is the atomic mass,
and~$\lambda(t)$ is the time-dependent control trajectory for translating the potential across space.
In practice,
evolution is solved for an initial state
$\psi_\text{in}:=\psi(x;0)$
numerically to obtain a final state
\begin{equation}
\label{eq:finalstate}
\psi_\text{fin}:=\psi_\text{fin}(x;T)
\end{equation}
at final time~$T$.
We take the following wave function normalization condition
\begin{equation}
\int\text{d}x\left|\psi(x;t)\right|^2=1.
\end{equation}
The energy of the BEC is a functional~\cite{dion2007},
\begin{align}
\label{eq:energy}
E[\psi](t)
=&\int\text{d}x\bigg[\frac{\hbar^2}{2m}\left|\partial_x\psi(t)\right|^2\nonumber\\
&+V(x)\left|\psi(t)\right|^2+\frac{g}{2}|\psi(t)|^4\bigg].
\end{align}
The energy of the BEC can be inferred from a series of time-of-flight images~\cite{bucker2011,van2014}.

We label the~$i^\text{th}$ excited state of the BEC by~$\phi_i$ with~$\phi_0$ representing the ground state.
Please refer to Appendix~\ref{appe:bec} for detailed definitions of the ground state and excited states of the BEC.
We label the ground state and the first excited state of the BEC as~$\ket0$ and~$\ket1$,
respectively.
We use
\begin{equation}
\label{eq:p_i}
p_i=\|\braket{\phi_i|\psi}\|^2
\end{equation}
to represent the probability for the atom to be found in the ground state ($i=0$),
and for each $i^\text{th}$ excited state for $i\in[3]$
where
\begin{equation}
\label{eq:[M]}
[M]:=\{1,\dots,M\}
\end{equation}
is a convenient set-theoretic notation.

According to~$\ket0$ and~$\ket1$,
we can define the following commonly used states
\begin{equation}
\label{eq:+-}
\ket\pm:=\ket0\pm\ket1,
\ket{\pm \text{i}}
    :=\ket0\pm\text{i}\ket1
\end{equation}
with normalization coefficients suppressed.

BEC states are complex-valued vectors and can be treated as vectors so a superposition of the ground and excited states is meaningful.
However, nonlinear evolution means that mapping the state forward in time fails to respect superpositions.
The fact that BEC evolution is not a linear operator means that designing nonlinear quantum gates,
e.g., for quantum computing,
is also complicated by the fact that these gates are not linear maps.

Now we discuss the function that should be used to describe the time-dependent potential $V$ in Eq.~(\ref{eq:Hdef}).
Properly controlling this time-dependent potential is vital for transforming the state of the BEC in the desired way.
Experimentally, the potential is used to tightly compress the atoms during the cooling process and it is also used to hold the BEC when the BEC is formed~\cite{ketterle1999}.
The potential used to trap the BEC can be realized on the atom chip~\cite{trinker2008},
and the displacement of the potential can be obtained by modulating radio-frequency currents~\cite{lesanovsky2006adiabatic,hofferberth2006radiofrequency}.
The potential along the shaking direction can be expressed in parametric form by a 6th-order polynomial
\begin{equation}
\label{eq:6thorderpotential}
 V(\bm{\alpha};x)/h
 =\frac{\alpha_2}{2}\left(\frac{x}{l}\right)^2+\alpha_4\left(\frac{x}{l}\right)^4+\alpha_6\left(\frac{x}{l}\right)^6,\,
 \bm\alpha\in\mathbb{R}^3,
\end{equation}
with length
\begin{equation}
\label{eq:length}
l:=\frac{\sqrt{h/(m\alpha_2)}}{2\pi}
\end{equation}
corresponding to the characteristic length of the harmonic part.
The shaking trajectory~$\lambda(t)$ as in~(\ref{eq:Hdef}) is used to control the transfer of the BEC from the ground state to the first vibrationally excited eigenstate~\cite{bucker2013}
and thereby to achieve Ramsey interferometer for motional states of the BEC~\cite{van2014}.

\subsection{Nonlinear quantum mechanics}
\label{subsec:nonqm}
We discuss the essential properties of nonlinear quantum mechanics.
We first review the presence of nonlinear quantum mechanics in some physical phenomena.
Then, we review the similarities and differences between nonlinear quantum mechanics and linear quantum mechanics.
Finally, we review the concept of unitary operators in nonlinear quantum mechanics.

Nonlinear quantum mechanics appears in macroscopic quantum systems,
such as superconductivity~\cite{cyrot1973ginzburg},
superfluidity~\cite{adhikari2008nonlinear},
and BECs~\cite{pitaevskii1961,gross1961}.
The macroscopic quantum effect results from the collective motion and excitation of particles under certain conditions,
such as extremely low temperature, high pressure, or high density.
Under such conditions, a huge number of microscopic particles condense,
resulting in a highly ordered and long-range coherent low-energy state~\cite{pang2005quantum}.
The theories describing and explaining these physical phenomena all involve nonlinear quantum mechanics,
such as Ginzburg–Landau theory~\cite{cyrot1973ginzburg},
used to explain the properties of superconductivity,
and the GPE~\cite{pitaevskii1961,gross1961} describes the dynamics of superflows and BECs at very low temperature.

We review the similarities and differences between nonlinear and linear quantum mechanics.
Their main similarities are~\cite{pang2005quantum}
(1) the system is described by a wave function;
(2) both have the concepts of operators and averages;
(3) both have the concepts of stationary state and eigenvalue;
and their main differences are~\cite{pang2005quantum}
(1) the absolute square of the wave function is no longer the probability for finding the microscopic particle at a given point in the space-time, but gives the mass density of the microscopic particles at that point;
(2) operators are no longer linear operators;
(3) the principle of superposition of states no longer holds.

We now review nonlinear unitary operators.
For nonlinear quantum mechanics,
unitarity is no longer inner-product preserving but rather just norm preserving~\cite{schwartz1997}.
Unitary, i.e., norm-preserving, evolution is generated by exponentiating the Hamiltonian,
but of course this unitary evolution is not necessarily a linear map.
Bringing nonlinearity into quantum information raises startling, subtle issues such as being able to distinguish between nonorthogonal states and to perform unstructured search~\cite{childs2016optimal}.

\subsection{Gate fidelity}
\label{subsec:gatefidelity}
We discuss relevant aspects of quantum gates in standard linear quantum mechanics.
First, we explain quantum gates and their matrix representations.
Then, we discuss the fidelity of quantum gates and the relevance of fidelity to assessing gate performance.

Now we discuss quantum gates and their unitary matrix representation for standard linear quantum mechanics.
In the computational basis,
corresponding to logical zero~$\ket0$and logical one~$\ket1$,
quantum gates can be expressed as unitary matrices.
Quantum gates acting on~$n$ qubits are represented by~$2^n\times 2^n$ unitary matrices.
A single-qubit unitary gate can be expressed as a~$2\times2$ complex matrix
\begin{equation}
\label{eq:gatematrix}
\bm{U}=\begin{pmatrix}
	u_{11} & u_{12} \\
       u_{21} & u_{22}
    \end{pmatrix},\;
    u_{ij}:=\langle i|U|j\rangle.
\end{equation}
In standard quantum mechanics,
a unitary operation is a linear isometry (preserves inner product).

We now explain state fidelity,
which quantifies how closely the resultant final state~(\ref{eq:finalstate})
approximates the desired target state~$\ket{\psi}_{\text{tar}}$.
The fidelity between the final and target state
(achieved by applying gate~$\bm U$
and by integrating the nonlinear evolution meant to approximate~$\bm U$,
respectively)
is expressed as
\begin{equation}
\label{eq:statefidelity}
\mathcal{F}_{\bm{U}}(\psi)
:=\left|{_\text{fin}}\braket{\psi|\psi}_{\text{tar}}\right|^2.
\end{equation}
If the target and final states have different Hilbert space dimension,
such as the case arising for target state being a qubit but final state having support beyond the qubit basis states,
we compute the fidelity~(\ref{eq:statefidelity})
over whichever of the two pertinent Hilbert spaces has the higher dimension.

Performance of a unitary gate is conveniently assessed by its average gate fidelity,
which averages state fidelity based on a uniform distribution of input states for the gate~\cite{nielsen2002}.
A subtlety with nonlinear quantum-gate fidelity is that the gate acting on a superposition of initial quantum states is not necessarily the superposition of the gate acting on the two initial quantum states.
Nevertheless,
we also assess the performance of nonlinear quantum gates by average gate fidelity,
whose definition survives the transition from linear to nonlinear quantum mechanics.
Specifically,
even for nonlinear quantum gates,
the average gate fidelity
for ideal gate~$\bm{U}$
is specified similarly as
\begin{equation}
\label{eq:averagefidelity}
\bar{\mathcal{F}}_{\bm{U}}
:=\int\text{d}\mu(\psi)\mathcal{F}_{\bm{U}}(\psi)
\end{equation}
for~d$\mu(\psi)$ the uniform, or Haar, measure with~$\psi$
labeling states
and the fideity integrand defined in Eq.~(\ref{eq:statefidelity}).
For subsequent convenience,
we employ the symbols~$\mathcal{F}_\text{max}$ and~$\mathcal{F}_\text{min}$ to represent the maximum and minimum fidelity, respectively.

\section{Approach }
\label{sec:approach}

In this section,
we discuss our approach to solving the problem of nonlinear quantum gates.
First we explain our model and our mathematical approach, building on the preliminaries explicated in~\S\ref{sec:background}.
We follow in the next subsection by explaining the searching and feasibility.
Finally, in the last subsection we explain our technique to find feasible potentials and shaking trajectories.
\subsection{Model and Mathematics}
\label{subsec:modelmath}

We describe the model used to implement nonlinear quantum gates on the BEC system.
We first explain the trapped atoms system and then follow in the next subsubsection by explaining the shaking method.

\subsubsection{Trapped atoms}

We follow the conventional scheme for trapped alkali atoms.
Here we explain our model for trapped rubidium.
Then we explain our proposal for how the nonlinear gate could be executed.
Finally, we discuss how resultant qubit should be measured.

The model we consider is the $^{87}$Rb BEC trapped in the one-dimensional potential as reviewed in~\S\ref{subsec:BECdynamics}.
The one-dimensional potential can be realized on an atom chip~\cite{reichel2011atom} and the shaking of potential can be realized by modulating radio-frequency currents~\cite{lesanovsky2006adiabatic,hofferberth2006radiofrequency}.
We assume temperature and interaction energy being smaller than the transverse trapping frequency; i.e., neither temperature nor interaction energy can excite the BEC.
In our simulation,
the potential is very anisotropic,
which is narrow in two dimensions and wide in one dimension.
The potential is in three dimensions and holds atoms in three dimensions but under special circumstances the dynamic of the BECs is one-dimensional,
that is the other degrees of freedom are not excited~\cite{van2014}.
The BEC will be excited to motional states due to the potential shaking.

Our nonlinear single-qubit gate employs the ground state and the first excited state of its motional degree of freedom of the BEC as the computational basis.
Nonlinear quantum gates are realized by finding feasible shaking trajectories of the potential,
where feasibility is established by a threshold condition for average gate fidelity~(\ref{eq:averagefidelity}).
We find feasible parameters for nonlinear evolution to realize the nonlinear gate by using optimization methods;
typically, control theory is discussed in the language of optimization,
which aims to find the best solution,
but, in practice,
feasibility problems are solved instead,
where feasibility is about the more modest problem of just finding a ``good enough'' solution~\cite{spiteri2018}.
Due to nonlinear evolution,
system evolution is a function of initial state, so we need to establish a single potential shaking trajectory that is sufficiently good based on averaging over all initial states.

We now describe the measurement of the BEC.
In our simulation,
we simulate the wavefunction of the BEC through solving the GPE.
In the experiment,
populations of the ground state and first excited state of the BEC are inferred from the evolution of the momentum density, which is obtained by time-of-flight images~\cite{van2014}.

\subsubsection{Shaking}

We now explain the trapping potential employed in our simulation.
We first explain the form of the potential and then we explain how the parameters of the potential relate to our nonlinear gate problem.
Finally, we explain how to evaluating candidates for potential parameters and candidates for a feasible shaking trajectory by sampling initial states uniformly on~$\mathcal{S}^2$.

We translate the potential~$V(x)$ according to the control trajectory~$\lambda(t)$,
without changing the shape of~$V$,
to realize the nonlinear quantum gate~\ref{eq:Hdef}.
We employ an anharmonic potential
this anharmonicity makes the energy level spacing different,
which can inhibit the transition of atoms to the high levels.
Making the potential strongly anharmonic
implies energy-level spacing is uneven, which suppresses leakage to higher energy levels.
In a harmonic potential, even at zero non-linearity,
a classical (coherent) drive produces a coherent state of excitation and not a number state (occupation of only the first excited state).
We truncate our even-order polynomial from sixth to fourth degree.
We neglect the sixth-order term
$\alpha_6$ in Eq.~(\ref{eq:6thorderpotential}) because we need to reduce the size of the search space to achieve computational tractability
and the fourth-order suffices to ensure unequal spacing between energy levels,
which we need for effective control.

The potential parameters $\alpha_{2,4}$~(\ref{eq:6thorderpotential})
are expressed briefly by $\bm\alpha\in\mathbb{R}^2$.
We restrict this domain to within an order of magnitude for the experimental choices of $\bm\alpha$~\cite{van2014}
and then choose a random~$\bm\alpha$
from this domain.
After choosing initial~$\bm\alpha$,
we commence with an initial trajectory, trajectory~$\lambda(t)$,
but our algorithm operates in the frequency domain, hence with the Fourier transform~$\tilde\lambda(f)$
with appropriate frequency bandwidth and discretizing the frequencies.
The search for a feasible trajectory is then executed over the frequency domain.

Evaluating a given choice of~$\bm\alpha$
and~$\tilde\lambda(f)$
requires evaluating $\bar{\mathcal{F}}_{\bm{U}}$~(\ref{eq:averagefidelity}) to assess gate fidelity
for a given gate~$\bm U$.
Thus, $\bar{\mathcal{F}}_{\bm{U}}$ is an integral over initial single-qubit states~(\ref{eq:statefidelity}).    
We convert this integral to a sum
by choosing an appropriate sampling over initial states
\begin{equation}
\label{eq:avfidsum}
\left\{\ket{\psi_i};
    i\in[M]\right\}
\end{equation}
with~$[M]$ defined in Eq.~(\ref{eq:[M]}).
For qubits,
this sampling is uniform from~$\mathcal{S}^2$.
From this sampling of initial states~(\ref{eq:avfidsum}),
we obtain the sum
\begin{equation}
\label{eq:avgatefidelity}
\bar{\mathcal{F}}_{\bm{U}}
=\frac1{M}\sum_{i=1}^M
\mathcal{F}_{\bm{U}}(\psi_i)
\end{equation}
with~$\mathcal{F}_{\bm U}(\psi_i)$
defined in Eq.~(\ref{eq:statefidelity}).
\subsection{Searching and feasibility}
\label{subsec:searchingfeasibility}

We now explain our searching strategy and pose the problem of finding appropriate potentials and shaking trajectories as a feasibility problem.
We begin by explaining our method for searching potential parameters and shaking trajectories.
Then we explain the feasibility framework of our problem.
Finally, we explain how state sampling is performed and how average gate fidelity is estimated.

\subsubsection{Searching shaking functions}

The average gate fidelity~(\ref{eq:avgatefidelity})
depends not only on the target~$\bm U$
but also on the potential parameters
$\bm\alpha$~(\ref{eq:6thorderpotential}),
after having chosen a shape for the family of potential functions.
This fidelity also depends on the choice of trajectory
$\lambda(t)$~(\ref{eq:Hdef}).
The trajectory function's Fourier transform is more convenient for the search.
We explain these issues in this subsubsection.

To make this variation over potential parameters and shaking trajectory clear and explicit,
we write the fidelity threshold condition as
\begin{equation}
\label{eq:avgatefidthreshold}
\bar{\mathcal{F}}_{\bm{U}}\
    \left(\bm\alpha,\tilde{\bm\lambda}\right)
    \geq\mathcal{F}_\text{thr},
\end{equation}
which is the feasibility condition for our search over potential parameters and shaking trajectories.
For a set of parameters~$\left(\bm\alpha,\tilde{\bm\lambda}\right)$, if Eq.~(\ref{eq:avgatefidthreshold}) holds, it is said that this set of parameters is feasible; otherwise it is not feasible.
Here~$\mathcal{F}_\text{thr}$
is a given average gate fidelity threshold value,
and
\begin{equation}
\label{eq:tildeomegadiscrete}
\tilde{\bm\lambda}
    :=\left(\tilde\lambda_i\right),\;
\tilde\lambda_i:=\tilde\lambda(f_i) \in\mathbb{C},
\end{equation}
is the Fourier transform of the shaking trajectory over a finite, discrete mesh $\{f_i\}$ of frequencies.

We proceed in two stages:
fix~$\bm\alpha$ and search for feasible (complex)~$\tilde{\bm\lambda}$
and then change~$\bm\alpha$
if a feasible~$\tilde{\bm\lambda}$ is not found,
i.e.,
\begin{equation}
\label{eq:fixsearch}
\text{search}(\bm\alpha)
\leftrightarrow
\text{fix}(\bm\alpha),\text{search}(\tilde{\bm\lambda}).
\end{equation}
If the feasibility condition is not met for any tested~$\tilde{\bm\lambda}$,
a new~$\tilde{\bm\lambda}$ is found and tested by whether its resultant average gate fidelity passes the threshold test~(\ref{eq:avgatefidthreshold}).
If feasibility is not met for any~$\tilde{\bm\lambda}$,
then a search is conducted for a new set of potential parameters~$\bm\alpha$
and then,
after selecting new~$\bm\alpha$,
we repeat the search for a feasible trajectory~$\tilde{\bm\lambda}$
for this new potential function.

After finding an feasible trajectory for the nonlinear Hadamard gate~(\ref{eq:Hgate}),
we employ this nonlinear Hadamard gate for the Ramsey interferometer,
which involves a nonlinear Hadamard gate followed by free evolution and then a second nonlinear Hadamard gate.
Although this second nonlinear Hadamard gate can differ from the first, we employ the same nonlinear Hadamard gate for both cases.

\subsubsection{Feasibility}

Now we explain how we find feasible values of~$\bm\alpha$
and~$\tilde{\bm\lambda}$
for the fidelity threshold test~(\ref{eq:avgatefidthreshold}).
Searches for~$\bm\alpha$ and~$\tilde{\bm\lambda}$
are conducted separately in an alternating fashion.
For fixed~$\bm\alpha$,
the search for~$\tilde{\bm\lambda}$ is performed by global optimization methods, and the search for~$\bm\alpha$
is a brute-force search, as we now explain.

Greedy searches~\cite{coleman1996interior} are fast but only reliable for convex optimization problems;
the alternative is global optimization,
which seeks the best solution over the entire parameter domain.
As our focus is on feasibility rather than the more ambitious task of optimizing,
we first seek a feasible solution locally using greedy optimization methods.
If greedy methods fail,
we turn to global optimization methods~\cite{ugray2007},
which are more computationally expensive but manage to find feasible solutions that are not local.
In our search for feasible~$\tilde{\bm\lambda}$,
we find that greedy searches fail so we resort instead to a global search algorithm.

If a feasible trajectory is found through searching $\tilde{\bm\lambda}$,
the search is finished and the feasible solution delivered.
If, on the other hand,
the search fails to deliver feasible~$\tilde{\bm\lambda}$,
then the potential is modified by searching for new potential parameters~$\bm\alpha$.
In the search process, for each~$\bm\alpha$,
we solve the optimal trajectories
$\{{\bm\lambda}^i;i\in[M]\}$,
with~$i$ denoting the $i^\text{th}$ element of the size-$M$ Fibonacci lattice using Hohenester's quantum-control method in OCTBEC,
as discussed in Appendix~\ref{appe:oct},
For each control trajectory~${\bm\lambda}^i$,
we require it to control the corresponding initial state~$\ket{\psi_i}$ (\ref{eq:avfidsum}) to evolve to the final state for a given gate with gate fidelity greater than 99.99\%.
If this condition is not met, we continue to search for the next~$\bm\alpha$.
For this search,
we employ a simple brute-force method for searching a new~$\bm\alpha$.
Specifically, we make a regular lattice
of dimension equal to the dimension of the vector~$\bm\alpha$
and proceed to the next neighbor.
This simple search technique has proven to be effective for this problem.

Assessing whether a given solution,
namely $(\bm\alpha,\tilde{\bm\lambda})$,
is feasible or not,
we compute a cost function
and determine whether this cost surpasses a threshold condition or not.
The cost function we use is average gate fidelity with the threshold condition given in Eq.~(\ref{eq:avgatefidthreshold}).
To estimate the cost function,
we sample uniformly over~$\mathcal{S}^2$
as explained in Eq.~(\ref{eq:avgatefidelity}).

\subsubsection{Sampling and fidelity estimate}
\label{subsubsec:sampling}
Sampling involves averaging over input states,
which involves a prior,
i.e., an initial distribution of states,
and we assume a uniform prior,
which in our simulation is achieved by choosing points in the Fibonacci lattice,
with this Fibonacci lattice a convenient way to sample efficiently the sphere in an unbiased way.
We explain relevant concepts of,
and the mathematical expression for,
the Fibonacci lattice.
Finally, we show the link between the Fibonacci lattice and the pure quantum state.

For efficient sampling,
we choose the set of points known as the Fibonacci lattice as these points are uniformly distributed on~$\mathcal{S}^2$ and have approximately isotropic
resolution~\cite{swinbank2006}.
The number of points we use is effectively a hyperparameter that we obtain by preliminary numerical testing,
and this hyperparameter is also chosen as a multiple of the number of core processors in the cluster, for convenience.
The Fibonacci lattice is a mathematical idealization of natural patterns arising in repeated
plant elements, such as the scales of pineapples~\cite{gonzalez2010},
and our Fibonacci lattice approach effectively approximates the average fidelity integral~(\ref{eq:averagefidelity}) while avoiding the inconvenience of random sampling techniques for choosing points on the sphere.

In our simulation,
the average fidelity of the gate~(\ref{eq:avgatefidelity})
is calculated by a sum of points on the Fibonacci lattice rather than integrating over a continuum of states on~$\mathcal{S}^2$,
which is computationally expensive.
We sample points
\begin{equation}
\label{eq:spherecoords}    
\left\{\bm{r}_i=(x_i,y_i,z_i)\in\mathcal{S}^2\subset\mathbb{R}^3;
i\in[M]\right\}.
\end{equation}
These points are coordinates
\begin{align}
\label{eq:fibonaccipoints}
x_i=&\sqrt{1-z_i^2}\cos(2\pi i\zeta),\nonumber\\
y_i=&\sqrt{1-z_i^2}\sin(2\pi i\zeta),\\
z_i=&\frac{2i-1}{M}-1,\,
\zeta:=\frac{\sqrt{5}-1}2\nonumber
\end{align}
of the Fibonacci lattice on~$\mathcal{S}^2$
with~$\zeta$ the golden-ratio conjugate~\cite{swinbank2006}.
These points~(\ref{eq:fibonaccipoints})
are parametrized in polar and azimuthal angles of $\mathcal{S}^2$
by
\begin{equation}
\label{eq:fibonacciangles}
\theta_i:=\arccos z_i, \varphi_i:=\arccos\nicefrac{y_i}{x_i},
\end{equation}
respectively,
and the corresponding single-qubit states are
\begin{equation}
\label{eq:singlequbitstate}
\ket{\theta,\varphi}
    :=\cos\nicefrac{\theta}{2}\ket0+\text{e}^{\text{i}\varphi}\sin\nicefrac{\theta}{2}\ket1
    =\bm{R}(\theta,\varphi)\ket0
\end{equation}
for
$R\in\text{SU}(2)/\text{U}(1)$ the qubit rotation operator
and~$\theta$ and~$\varphi$
the Fibonacci angles~(\ref{eq:singlequbitstate}).
These Fibonacci-lattice points~\cite{swinbank2006},
are fairly evenly distributed on the Bloch sphere~$\mathcal{S}^2$.

\subsection{Methods}

Now that we have explained our model and the pertinent mathematics in our approach, here we elaborate on key methods that we employ.
Specifically,
we discuss our approach to three methods.
The first method concerns how we integrate the nonlinear differential equation.
The second method is about how we replace integration by sampling the Fibonacci lattice.
Finally, we discuss our methods for searching for a feasible trajectory and parameters for the quartic potential.

Our approach to integrating the nonlinear differential equation for evolution~(\ref{eq:1dham})
is solved using the OCTBEC toolbox~\cite{hohenester2014},
which we describe in~Appendix~\ref{appe:oct}.
We first define a position grid, potential parameters, a discrete time step~$\Delta t:=T/(n-1)$ for~$n$ time steps,
an initial two-level state supported over the ground and first-excited motional modes,
and the nonlinear Hamiltonian,
and then we run the program for fixed~$T$.
The output is the complex wave function of the BEC over the discrete position grid at time~$T$.

We sample over initial states~(\ref{eq:singlequbitstate}) corresponding to Fibonacci points~(\ref{eq:fibonaccipoints}).
As our cluster has 24 cores,
we choose 24 Fibonacci states for efficiently sampling average gate fidelity.
Although the input is a single-qubit state,
i.e., a pure state of a two-level system,
the output could be a multilevel state,
which we truncate successfully to four levels.

Each candidate shaking trajectory~$\tilde{\bm\lambda}$
has support over a wide range of frequencies,
which makes searching for optimal trajectories computationally expensive.
We devise a method for restricting the frequency domain for the search as we now explain.
First we solve the optimal trajectories
$\{{\bm\lambda}^k;k\in[M]\}$,
with~$k$ denoting the $k^\text{th}$ of the size-$M$ Fibonacci lattice using Hohenester's quantum-control method in OCTBEC,
as discussed in~Appendix~\ref{appe:oct},
plus the discrete Fourier transform.
First we define a spectral-averaging function
\begin{equation}
\label{eq:specav}
\Lambda_i
=\frac1{M}\sum_{k=1}^M
\left|\tilde{\lambda}^k_i\right|
\end{equation}
for the $i^\text{th}$ frequency component
$f_i$~(\ref{eq:tildeomegadiscrete}).
We then determine a connected subdomain~$I$ of frequencies for which~$\Lambda_i$ exceeds a cut off condition.
In other words we introduce lower- and upper-frequency cut offs by restricting~$i$
to~$I$ and thereby reduce computational overhead in searching for feasible trajectories for the general case.

Now that we have restricted the frequency domain to~$I$,
searching for a feasible~$\tilde{\bm\lambda}$,
denoted~$\tilde{\bm\lambda}_\text{feas}$,
based on average gate fidelity~(\ref{eq:avgatefidelity}),
corresponds to searching for a feasible vector over an $|I|$-dimensional vector space,
for~$|I|$ the cardinality of the restricted frequency mesh.
We commence the search with an initial trajectory candidate that has equal support over all frequency components.
Then we execute a search to find superior frequency-domain trajectories restricted to~$I$.
This procedure is terminated when the average gate fidelity exceeds the threshold value~(\ref{eq:avgatefidthreshold}),
and this frequency-domain description of a feasible trajectory is returned as the algorithmic output as well as the resultant fidelity.
If, on the other hand, a feasible trajectory is not found,
the algorithm searches for new parameters of the quartic potential and seeks a feasible trajectory in that case.
When we have obtained a feasible frequency distribution,
we use the inverse discrete Fourier transform to obtain the control trajectory in the time domain.
We can choose a different number of time domain samples to obtain a few control points but a steep control curve or more control points but a continuous control curve.
The sampling theorem sets the minimum number of sampling points.

The feasible trajectory~$\tilde{\bm\lambda}_\text{feas}$ has been obtained over~$M$ points of the Fibonacci lattice.
Due to nonlinearity of the evolution,
we do not presuppose that this trajectory is feasible over points outside this lattice.
Thus, we next characterize the fidelity for this candidate feasible trajectory over many points on~$\mathcal{S}^2$.
Our characterization is a ``heat map,''
i.e., a visualization of the data in two dimensions with color representing amplitude,
obtained by the Eckert~IV projection~\cite{snyder1989album},
which involves partitioning~$\mathcal{S}^2$ into small pieces.
The Eckert~IV projection method corresponds to an equal-area pseudo-cylindrical projection.
For this projection,
we choose a fine graining corresponding to a ``square degree'':
\begin{equation}
\label{eq:squaredegree}
64,800\text{ deg}^2=360^\circ\times180^\circ
\end{equation}
 elements corresponding to one degree steps both latitudinally and longitudinally on~$\mathcal{S}^2$.
 
Our search can be executed in a greedy way,
which searches locally only,
or globally,
which is typically much more computationally expensive but circumvents local traps that stymie greedy algorithms from finding global optima.
Greedy algorithms are excellent for convex optimization or for feasibility problems with adequate local optima.
We discover from our simulations that greedy algorithms are ineffectual for our nonlinear quantum gate problem,
at least for searching trajectories.
In our simulation,
when the potential parameters are determined,
we use MATLAB\textsuperscript{\textregistered}'s GlobalSearch solver~\cite{ugray2007} to find feasible trajectories.
Regarding the search for new parameters of the quartic potential,
we discover that a systematic mesh search is successful,
with the mesh search being deterministically stepping to nearest neighbours for $\bm\alpha\in\mathbb{R}^2$.

\begin{figure}
    \centering
    \includegraphics[width=0.8\columnwidth]{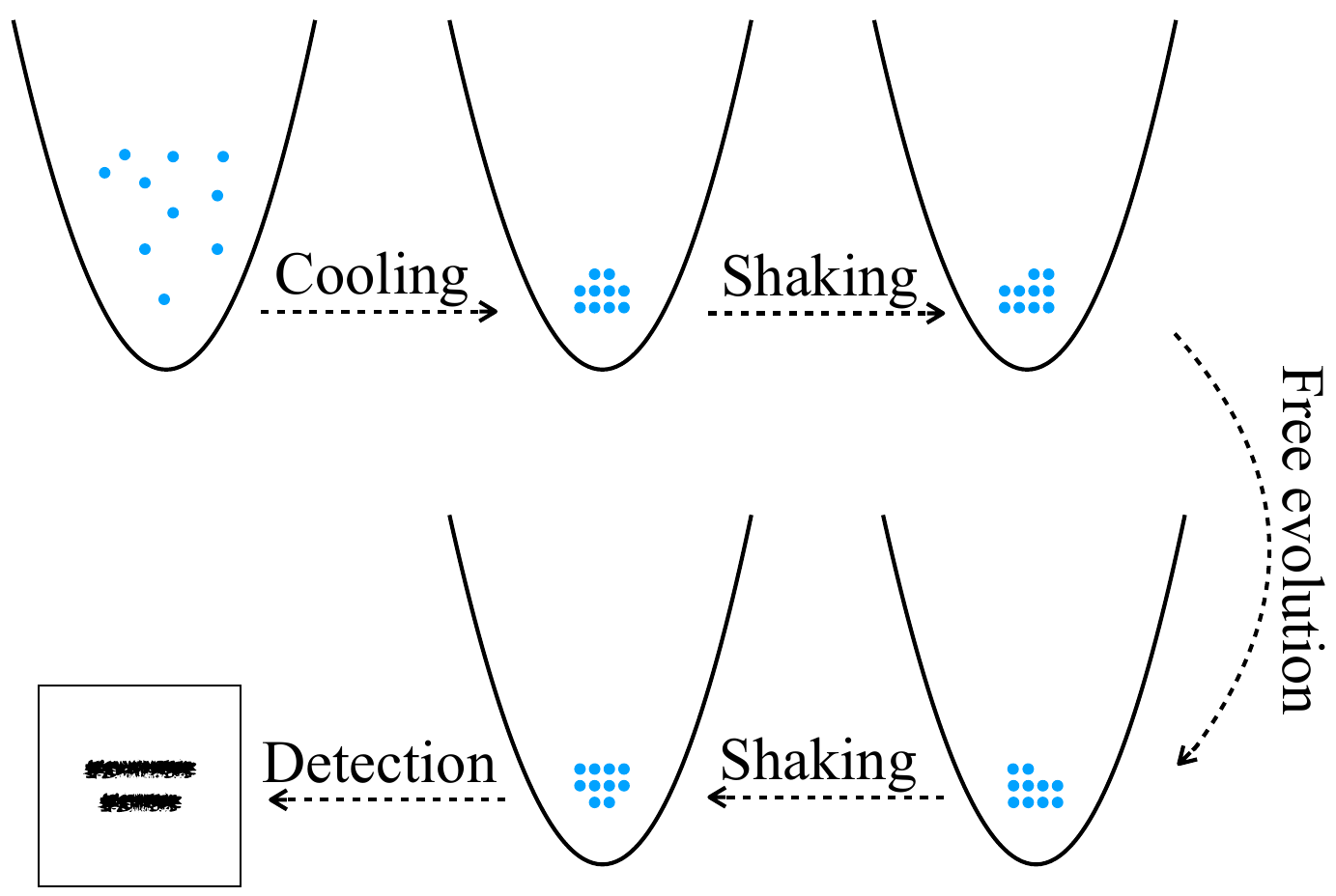}
    \caption{We show the six steps of Ramsey interferometry as (a)~trapped BEC, (b)~laser-cooling the BEC to a ground state,
    (c)~shaking the BEC to effect the initial Hadamard gate,
    (d)~letting the BEC undergo free evolution,
    (e)~shaking the BEC again,
    and (f)~finally imaging the BEC to ascertain the wave function distribution in position.}
    \label{fig:ramseyintsteps}
\end{figure}

We now show how to simulate Ramsey interferometry,
which uses the trajectory found by global search.
The sequence of steps for Ramsey interferometry is depicted in Fig.~\ref{fig:ramseyintsteps}.
First, the BEC is prepared in the ground state,
and then the BEC is shaken according to the feasible trajectory that we found;
this shaking transforms the BEC by a nonlinear Hadamard gate.
The time-dependent BEC wave function
during this shaking can be obtained by the Crank-Nicolson method,
which is incorporated into OCTBEC, a MATLAB\textsuperscript{\textregistered} toolbox.
Subsequent to the cessation of this shaking,
the BEC undergoes free evolution with the fixed potential being fixed (not shaking), and the wave function after free evolution can be obtained by OCTBEC.
This free-evolution step is followed by another Hadamard gate,
which we impose by employing the same feasible trajectory as for the first Hadamard gate.
Finally, the resultant wave function can be obtained, as an output of OCTBEC,
and plotting the resultant wave function is valuable as a check on our calculation.

\section{Results }
\label{sec:results}

We now present our results in this section.
According to our alternating two-stage process~(\ref{eq:fixsearch}),
first we search potential parameters
being employed to realize the Hadamard gate via shaking.
Then we present the results obtained by using the global search algorithm to find feasible control trajectories in the frequency domain.
Finally, we present our results from using the nonlinear Hadamard gate plus nonlinear free evolution to simulate Ramsey interferometry.

\subsection{Potential parameters}

In this subsection,
we present our results for feasible potential parameters.
First we describe our numerical search for feasible potential parameters,
according to Eq.~(\ref{eq:fixsearch}),
and then we present the potential parameters obtained from effecting this search.
Finally, we show,
for this resultant potential,
the ground and first excited state energies of the BEC for the cases that the nonlinear coefficient is zero and non-zero.

The goal is that in the case of a single state, we can use the optimal control trajectory obtained by the OCTBEC toolbox to control the BEC to reach the target state in the potential under this set of parameters.
If under this set of potential parameters, for the control of a single state, the state fidelity cannot reach a certain threshold, for example, 99.99\%, then it is impossible to further control all states to achieve quantum gates; therefore, this set of parameters is not feasible.
On the other hand,
if for~$M$ different states,
we use the OCTBEC toolbox to obtain~$M$ optimal shaking trajectories for controlling the BEC from initial state to target state.
After obtaining~$M$ optimal shaking trajectories,
we can get~$M$ final states by solving the differential equation~(\ref{eq:1dham}),
and then use Eq.~(\ref{eq:statefidelity}) to evaluate the feasibility of this guessing potential field.
If $\bar{\mathcal{F}}_{\bm{U}}$ is greater than 99.99\%,
we say this set of potential parameters is feasible.

It should be noted that feasible here means that for this set of potential parameters, for different initial states, we can use OCTBEC to obtain different optimal control trajectories, so that BEC reaches the corresponding final state. The nonlinear quantum gate requires that a unique control trajectory be found for different initial states which is a more difficult feasibility problem.

We now describe our search for quartic-potential feasible parameters~(\ref{eq:6thorderpotential}).
Our mesh search for feasible potential parameters~$\bm\alpha\in\mathbb{R}^2$
is restricted to the domain~$\alpha_2\in[500, 3000]$~Hz and
$\alpha_4\in[50, 8000]$~Hz with mesh step size 1~Hz.
The effective one-dimensional nonlinear coefficient~($\ref{eq:Hdef}$) in our simulation is
\begin{equation}
\label{eq:223}
g=h\times 223~\text{Hz}\,\mu\text{m}.
\end{equation}
Here we simulate the Hadamard gate~(\ref{eq:Hgate}) with integration time set to 1~ms.
The control trajectory~$\lambda(t)$ is discretized into 100 parts with time step~$\Delta t=0.01$~ms and the total number of samples~$n=101$.
The spatial range of the potential is limited to
\begin{equation}
x\in[-1.5, 1.5]~\mu\text{m}
\end{equation}
with space-step size set to 0.03~$\mu$m.

\begin{figure}
\includegraphics[width=0.8\columnwidth]{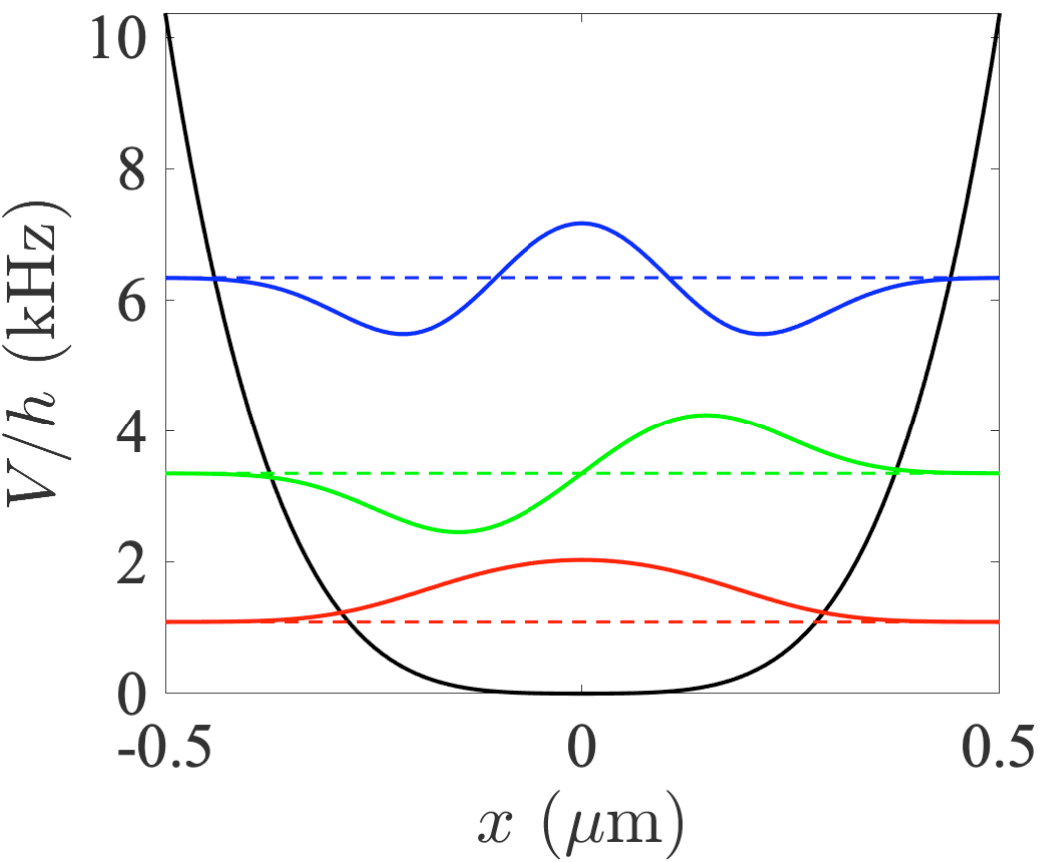}
\caption{%
	 Black line represents the shape of the potential for~$\alpha_2=533~\text{Hz}$ and $\alpha_4=7648\ \text{Hz}$.
    		The three horizontal dashed lines show the energy of the ground state and the first and second excited states of the BEC in this potential with nonlinear coefficient~$g=h\times 223$~Hz~$\mu$m, respectively.
		The wave functions corresponding to these energy levels are also plotted on the graph.
}
\label{fig:potential}
\end{figure}

Our initial candidate for potential parameters is~$\bm\alpha=(500, 50)$~Hz,
i.e., $\alpha_2=500~\text{Hz}$
and $\alpha_4=50~\text{Hz}$.
Then we execute a global search in the frequency domain with fidelity threshold~(\ref{eq:avgatefidthreshold}) equal to~99\% to find a feasible control trajectory to yield a Hadamard gate~(\ref{eq:Hgate}).
We obtain finally feasible trap parameters
\begin{equation}
\label{eq:vsigmal}
    \alpha_2=533~\text{Hz},\,
    \alpha_4=7648\ \text{Hz},
\end{equation}
which means that the algorithm selects a very anharmonious potential with the smaller~$\alpha_2$ and the larger~$\alpha_4$.
This set of parameters makes the potential very flat in the middle and steep on both sides.
It should be pointed out that the optimal parameters~$\alpha_2$ and~$\alpha_4$ obtained by the global search are close to the boundary, but we have obtained a feasible solution.
Figure~\ref{fig:potential} shows the shape of the potential under this set of parameters.

We calculate the ground and first excited states of the BEC using a function provided by OCTBEC,
which is based on an optimal damping algorithm~\cite{dion2007ground}.
For nonlinear coefficient~$g=0$,
and for the applicable parameters,
the initial degeneracy of the level
spacings is lifted with resultant energies
\begin{equation}
[\nu_0,\nu_1,\nu_2]=[0.90,3.18,6.19]~\text{kHz},\,
\nu_i:=E_i/\text{h},
\end{equation}
using the notation of Eqs.~(\ref{eq:E0}) and~(\ref{eq:E1}).
Relevant level spacings are given by
\begin{equation}
\nu_1-\nu_0=2.28~\text{kHz},\,
\nu_2-\nu_1=3.01~\text{kHz}.
\end{equation}
For nonlinear coefficient~(\ref{eq:223}),
the energies are
\begin{equation}
[\nu_0,\nu_1,\nu_2]=[1.09,3.35,6.34]~\text{kHz}
\end{equation}
and relevant level spacings are given by
\begin{equation}
\nu_1-\nu_0=2.26~\text{kHz},\,
\nu_2-\nu_1=2.99~\text{kHz}.
\end{equation}
The energy of the ground state and the first and second excited state are depicted in Fig.~\ref{fig:potential} for nonlinear coefficient~(\ref{eq:223}).

\subsection{Discrete fourier transforms of trajectories}
\label{subsec:Htrajectories}

We now show our results for frequency truncation.
We first show the frequency distribution that plays a role in the control process.
Then we present a feasible control trajectory obtained by a standard global search algorithm performed after truncating the frequency, i.e., restricting the bandwidth.
Finally, we show the effect of our feasible trajectory from the perspective of the BEC density distribution and the proportion of BEC population at different energy levels.

\begin{figure}
\includegraphics[width=0.8\columnwidth]{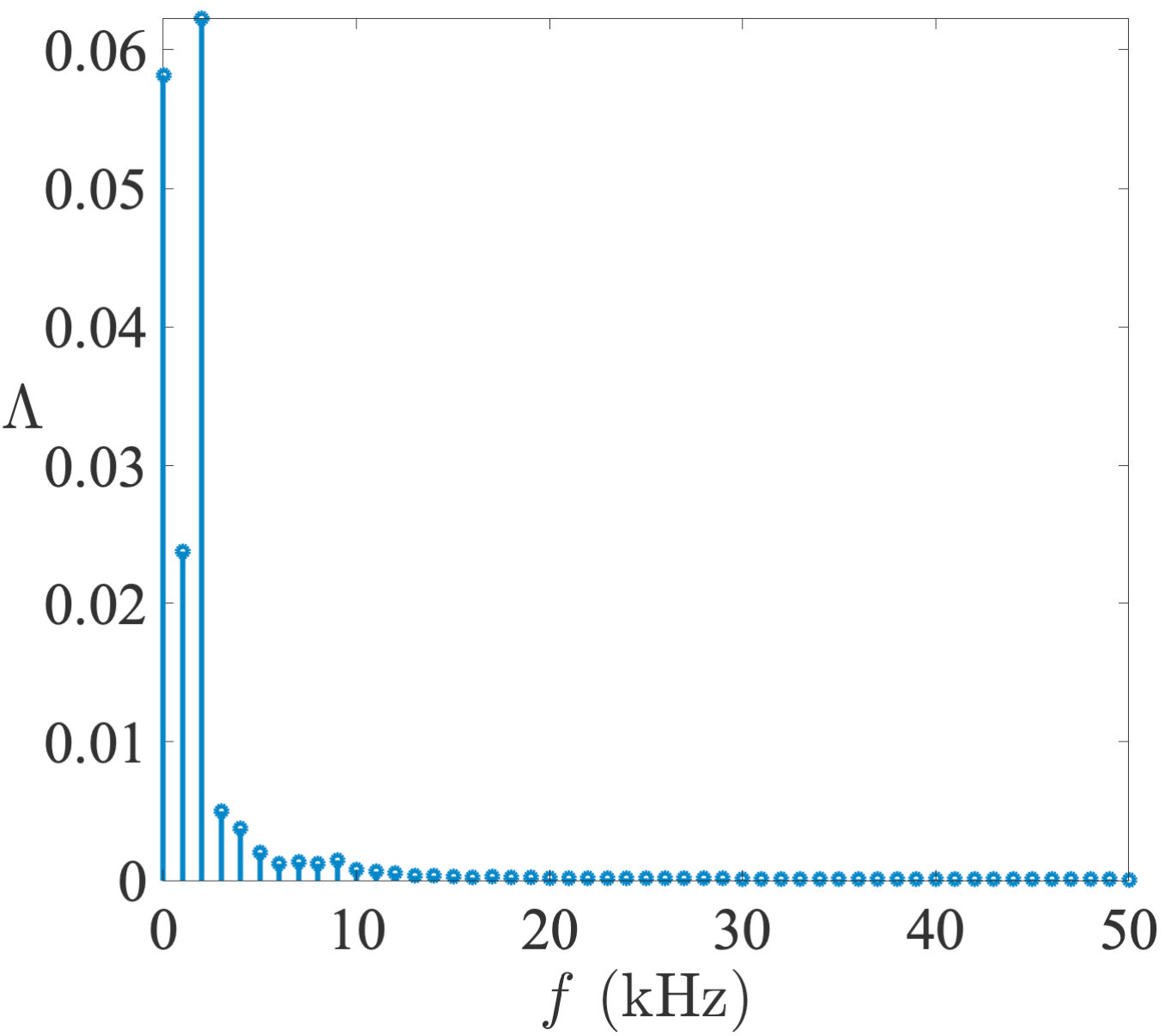}
\caption{%
	Spectral-averaging function~$\{\Lambda_i\}$
	vs frequency
	for $\{\bm\lambda^k;k\in[24]\}$ with~$[24]$
	defined in Eq.~(\ref{eq:[M]}). 
	Here $T=1$~ms,
	$\Delta t=0.01$~ms
	and $g=h\times 223$~Hz~$\mu$m.%
}
\label{fig:domainfrequency}
\end{figure}

In Fig.~\ref{fig:domainfrequency},
we depict the resultant spectral averaging function~(\ref{eq:specav}).
From this figure, we observe that
few low-frequency components play a significant role in
(i.e., support)
BEC control.
We neglect the zero-frequency, or DC case,
as it only translates the BEC in space,
which is not important here.
The dominant non zero frequency is $\nu=2$~kHz,
which can be explained by this~$\nu$ being closest to the frequency difference between ground and first excited states,
i.e.,
an energy difference of
$\nu_1-\nu_0=2.26$~kHz.

Although one frequency, namely, 2~kHz,
dominates the spectral averaging function,
a single frequency is insufficient to control the system well.
Fortunately,
we can choose a frequency range according to our requirements.
After selecting an appropriate frequency range,
global search is executed to obtain candidate control trajectories systematically.

\begin{figure}
\includegraphics[width=0.8\columnwidth]{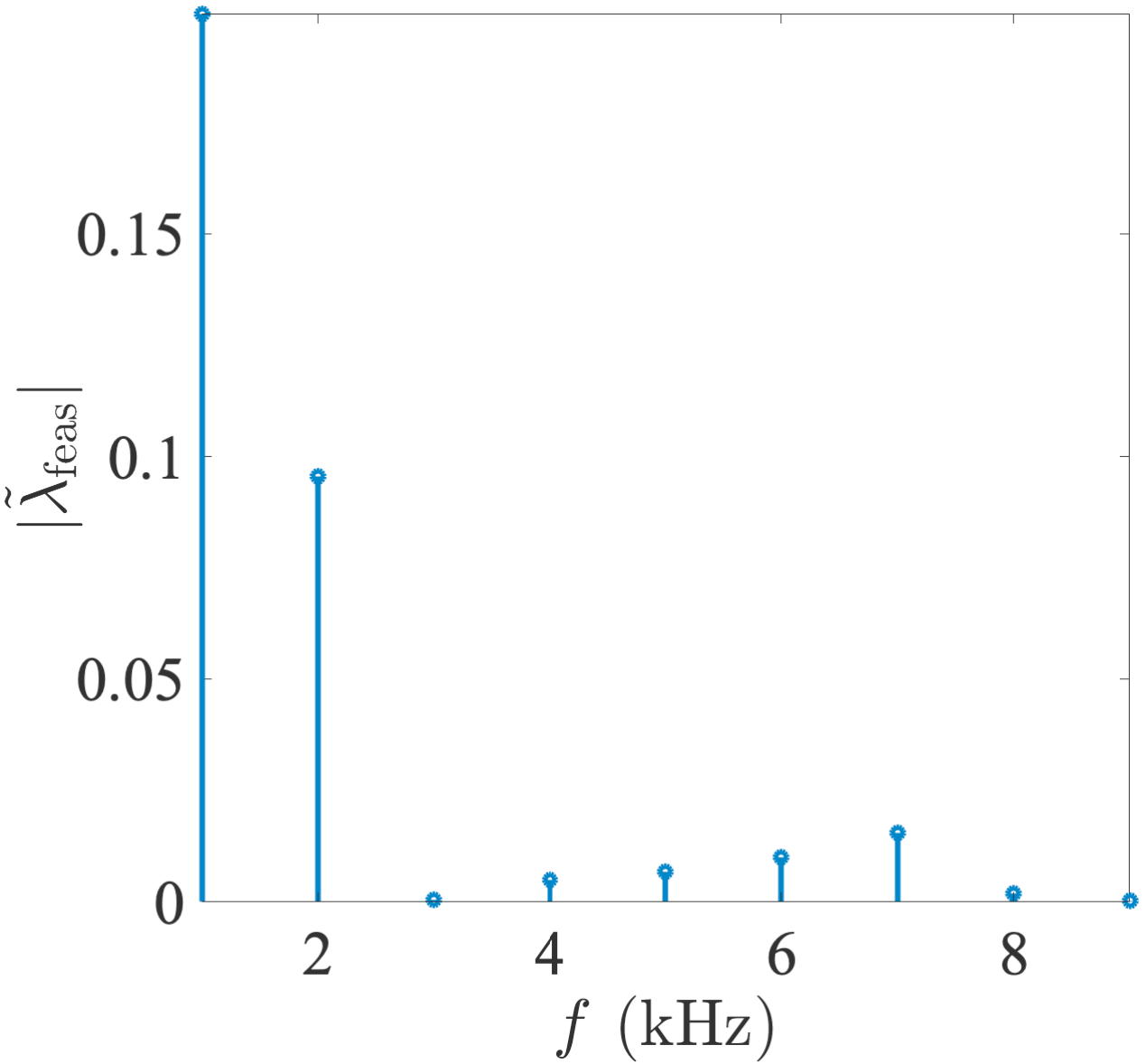}
\includegraphics[width=0.8\columnwidth]{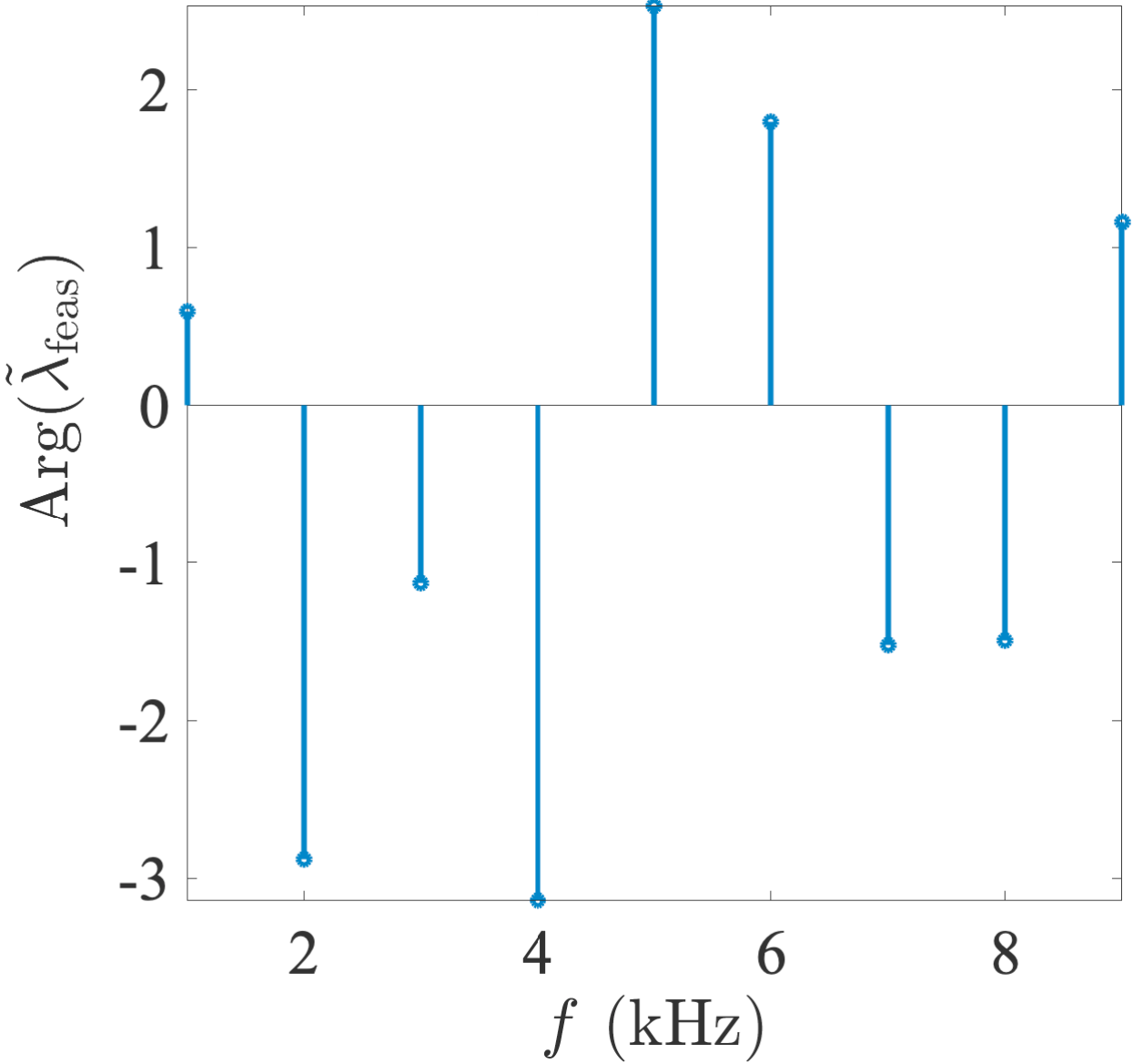}
\caption{%
	The modulus length (top) and phases (bottom) of~$\tilde{\bm\lambda}_\text{feas}$ with only nine frequency components being considered.}
\label{fig:frequencycutoff}
\end{figure}

\begin{figure}
\includegraphics[width=0.9\columnwidth]{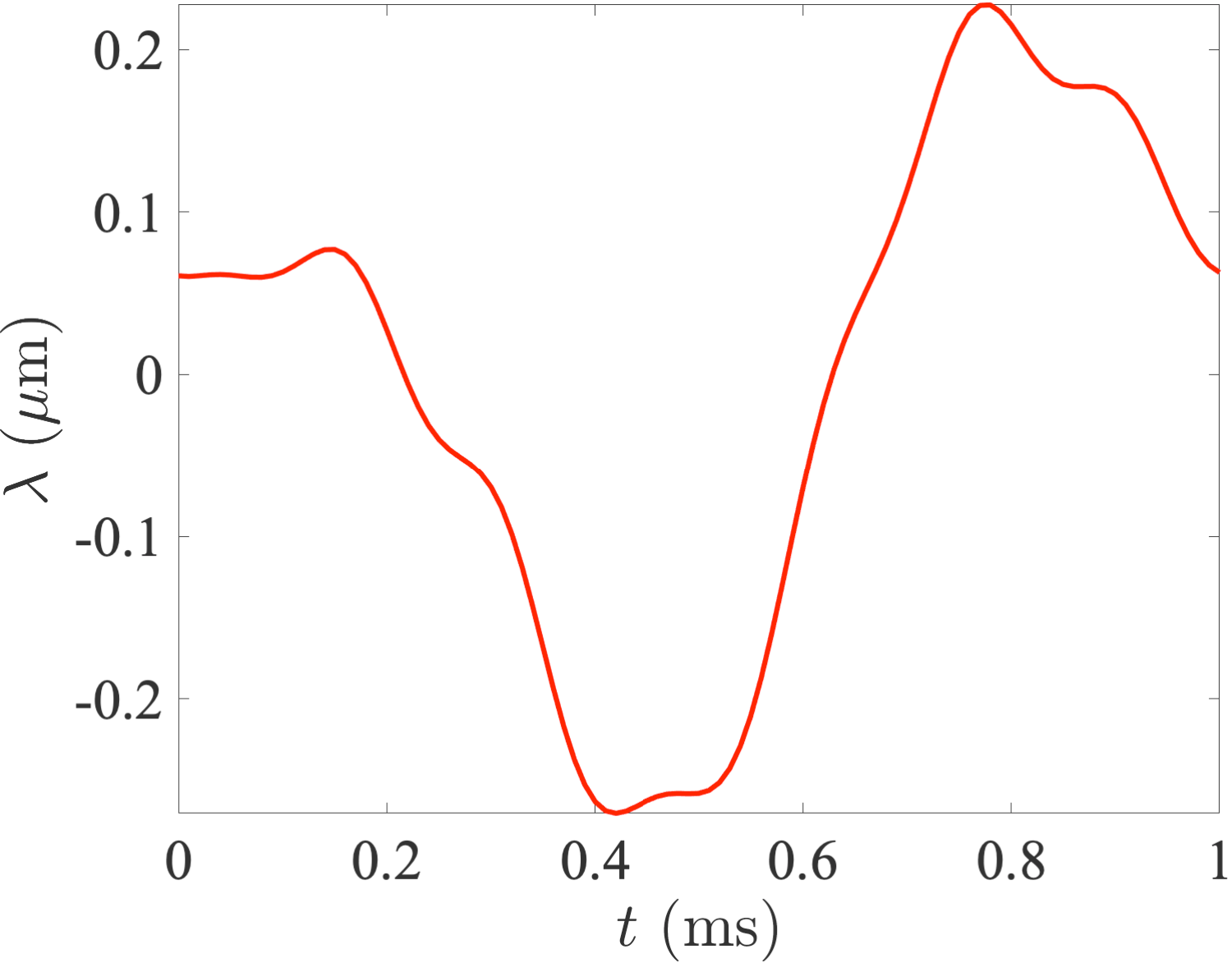}
\caption{%
	A time-dependent trajectory~$\lambda$,
	which effects a nonlinear Hadamard gate,
	for the restricted case that only nine frequency components are considered.%
}
\label{fig:timedomain}
\end{figure}

\begin{figure}
\includegraphics[width=0.85\columnwidth]{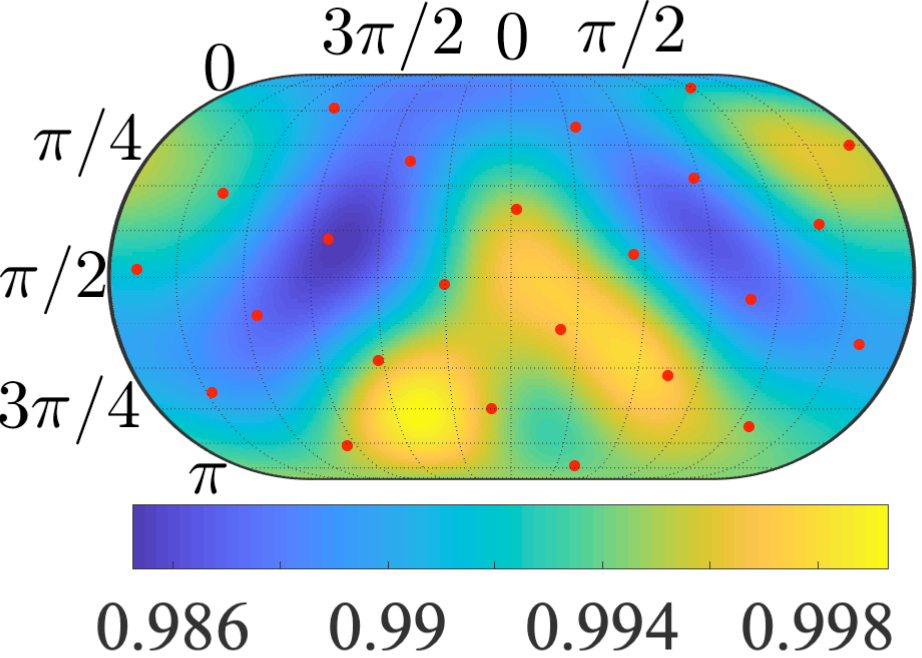}
\caption{%
	Gate fidelity for 64,800 initial states subject to evolution given by the Hadamard gate translational trajectory in Fig.~\ref{fig:timedomain} red line.
	The 24 points (red) used to obtain an optimal control trajectory by OCTBEC.
	The graphics are drawn using the Eckert IV projection.
}
\label{fig:fidelitydistribution}
\end{figure}

Figure~\ref{fig:frequencycutoff} shows,
via plotting~$\left|\tilde{\bm\lambda}_\text{feas}\right|$ and~$\text{Arg}\left(\tilde{\bm\lambda}_\text{feas}\right)$ vs frequency,
the nine frequency values needed to obtain a feasible trajectory~$\tilde{\bm\lambda}_\text{feas}$.
We see two dominant nonzero frequencies~$\{1,2\}$~kHz.
A plot of the corresponding time domain is shown in Fig.~\ref{fig:timedomain}.
This trajectory has been obtained,
similarly to our three-frequency case explained above,
by employing MATLAB\textsuperscript{\textregistered}'s GlobalSearch solver acting on the same $M=24$ Fibonacci points.
We plot the gate fidelity for 64,800 different initial states on~$\mathcal{S}^2$ as shown in Fig.~\ref{fig:fidelitydistribution}
to show how good our gate is for any initial state.
These results show that
\begin{equation}
    \mathcal{F}_\text{max}=99.93\%,\,\mathcal{F}_\text{min}=98.53\%,\,\bar{\mathcal{F}}=99.21\%.
\end{equation}
High fidelity might be achieved by using more frequencies.
We expect that incorporating more frequencies will yield a marginal gain of high fidelity as high frequency will cause BEC to be excited to a higher energy level.

\begin{figure}
\includegraphics[width=0.9\columnwidth]{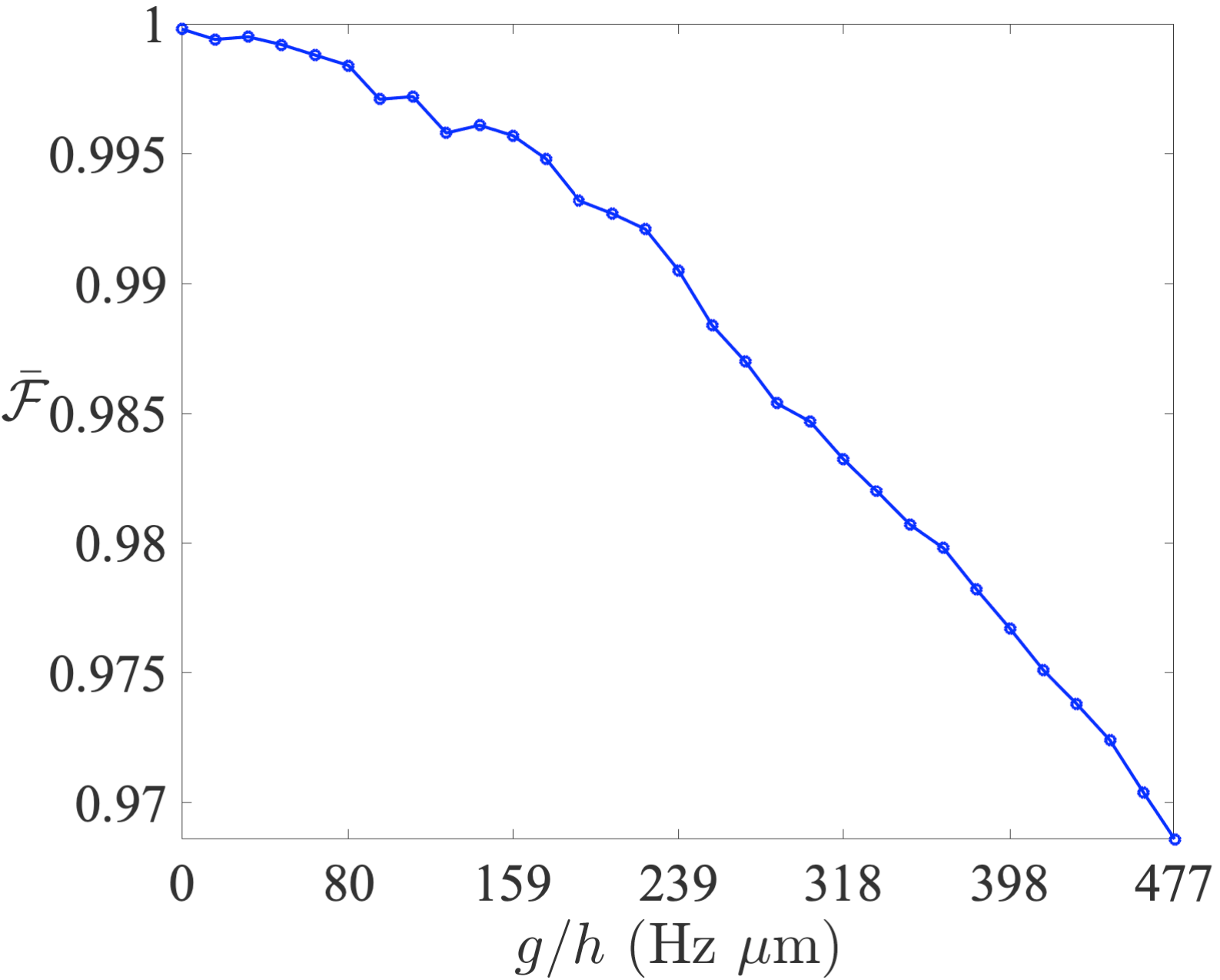}
\caption{%
	Average gate fidelity for different nonlinear coefficients~$g$.
	The average fidelity corresponds to the average of 64,800 corresponding states of~$\mathcal{S}^2$.
}
\label{fig:diffg}
\end{figure}

We also explored the best average fidelity that the global search algorithm can obtain under different nonlinear coefficients~$g$ with respect to potential parameters~(\ref{eq:vsigmal}).
As shown in Fig.~\ref{fig:diffg}, we simulated the situation from~$g=0$ to $g=h\times 477~\text{Hz}\,\mu\text{m}$.
We can estimate roughly the number of atoms in a BEC corresponding to the assigned nonlinear coefficient.
For~$a_\text{r}=900$ a.u.,
$g=h\times 477~\text{Hz}\,\mu\text{m}$ corresponds to about 980 atoms in the BEC.

\begin{figure}
\includegraphics[width=0.9\columnwidth]{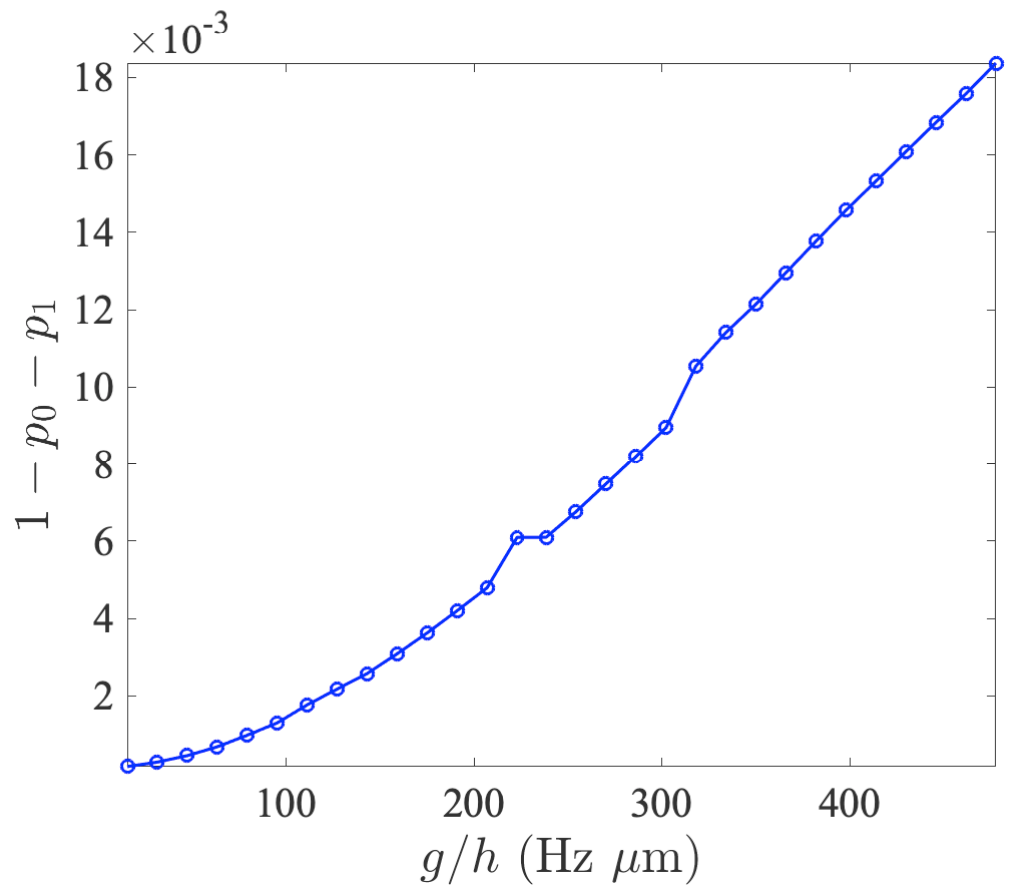}
\caption{%
	BEC leaking to the second and higher excited states for different nonlinear coefficients~$g$.
	The leaking corresponds to the average of 1,200 corresponding states of  Fibonacci points.
}
\label{fig:hightlevel}
\end{figure}

As the nonlinear coefficient increases, the average fidelity of the quantum gate decreases.
The average fidelity in this figure corresponds to the average of 64,800 corresponding states~(\ref{eq:squaredegree}) of~$\mathcal{S}^2$.
To explore why the average fidelity decays with increasing nonlinear coefficients,
we apply those optimal control trajectories to the quantum states corresponding to 1200 Fibonacci points; Fig.~\ref{fig:hightlevel} plots the probability of the BEC leaking to the second and higher excited states (average leakage over 1200 states).
Comparing Figs.~\ref{fig:diffg} and~\ref{fig:hightlevel}, we can see that the average fidelity degradation is due to the leakage of the BEC to higher energy levels.

\begin{figure}
\includegraphics[width=0.9\columnwidth]{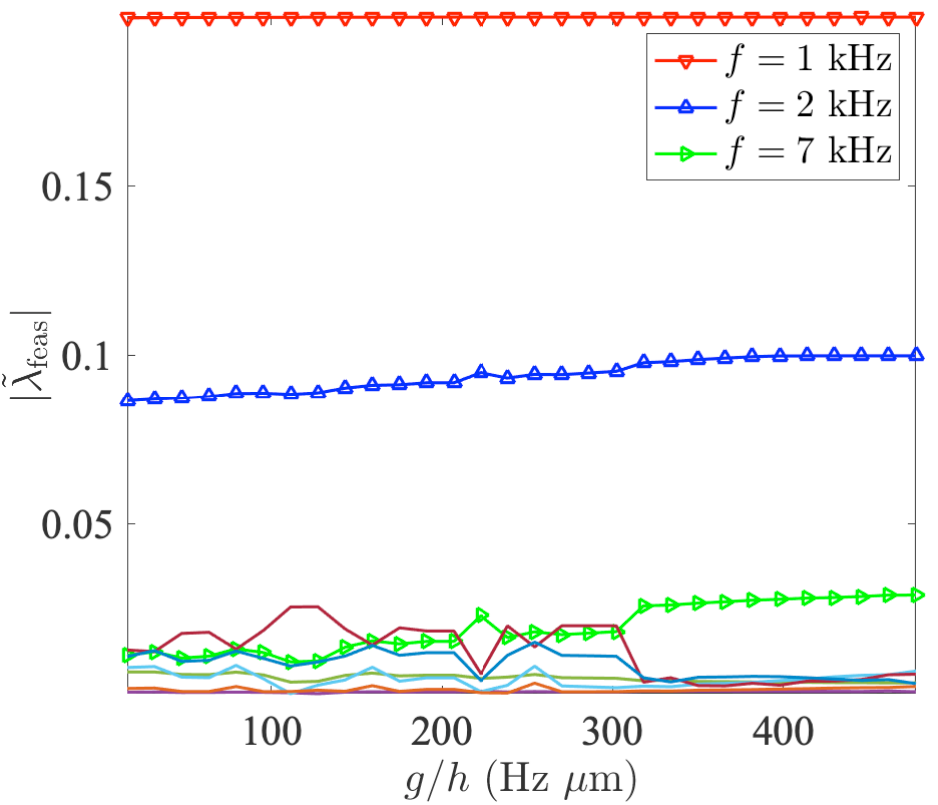}
\caption{%
	Modulus length of frequency components of the optimal control trajectory for different nonlinear coefficients~$g$.
	Only \{1, 2, 7\} kHz with obvious regularity are marked, and the rest of the frequencies \{3,4,5,6,8,9\} kHz are not marked.
}
\label{fig:g_with_f}
\end{figure}

In Figure~\ref{fig:g_with_f}, we plot the nonlinear coefficients~$g$ vs modulus length of frequency components~$\left|\tilde{\bm\lambda}_\text{feas}\right|$ contained in the optimal control trajectory with respect to the potential parameters~(\ref{eq:vsigmal}).
We can see that the proportion of frequency 1 kHz remains basically unchanged, the proportion of frequency \{2,7\} kHz increases slowly with the increase of the nonlinear coefficient, while the changes of other frequency components \{3,4,5,6,8,9\} kHz do not show obvious characteristics.

\begin{figure}
\includegraphics[width=0.8\columnwidth]{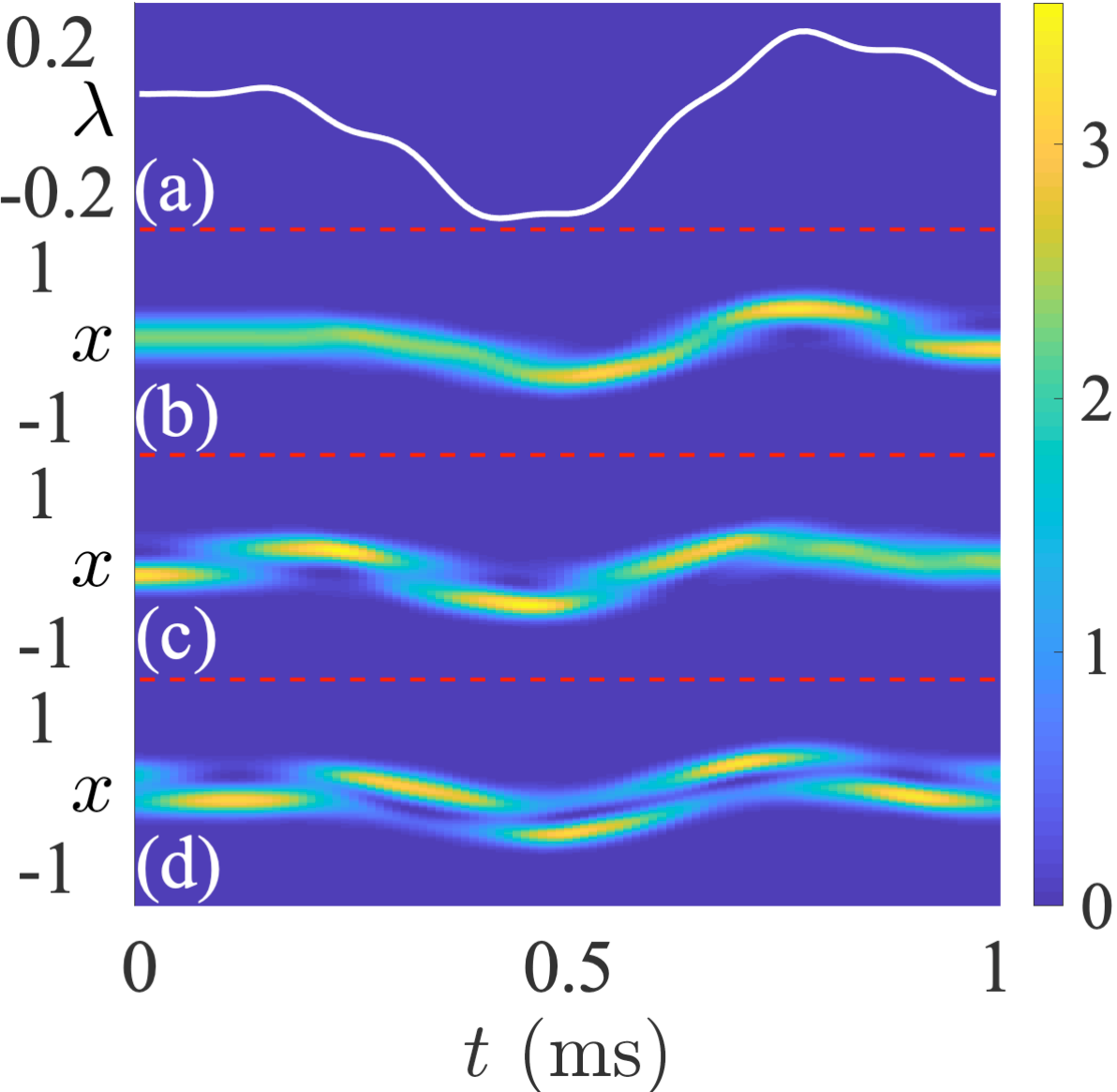}
\caption{%
	(a) Optimal control trajectory~$\lambda(t)$,
	over a 24-point Fibonacci lattice on~$\mathcal{S}^2$,
	for translating the potential (in~$\mu$m)
	from -0.1910~$\mu$m to 0.1608~$\mu$m
	for a Hadamard gate executed over 1~ms
	with time steps of 0.01~ms,
	and nonlinear coefficient~$g=h\times 223$~Hz~$\mu$m. 
	The frequency range is in [1, 9]~kHz.
	Population distribution~$\|\psi(x;t)\|^2$ vs time~$t$ and position~$x$ for different initial states,
	(b) $\ket0$, (c) $\ket+$, (d) $\ket{\text{i}}$, with gate fidelity 99.49\%,
	99.40\%, 98.28\%, respectively.
	Different colors represent the density as given in the legend in the right of the BEC at different~$x$.
}
\label{fig:densitydistribution}
\end{figure}

In order to show the effect of the Hadamard-gate control trajectory obtained this way,
we plot the position distribution probability for the BEC,
i.e., the BEC density,
during processing by shaking.
Figure~\ref{fig:densitydistribution}(a) depicts the feasible control trajectory we obtain through global search.
As Fig.~\ref{fig:densitydistribution}(a) is the inverse Fourier transform of Fig.~\ref{fig:frequencycutoff},
we see in Fig.~\ref{fig:densitydistribution}(a)
a trajectory dominated by a quasi sinusoidal curve with period~$1$~ms,
which is commensurate with the frequency 1~kHz.
The first-order correction to this sinusoidal modulation is the second frequency 2~kHz.
In Fig.~\ref{fig:densitydistribution}(b,c,d), we show~$\|\psi(x;t)\|^2$
during shaking for initial states~$\ket0$, $\ket+$, and~$\ket{\text{i}}$, respectively.
We see that the BEC oscillates back and forth closely linked to the control trajectory in~(a).
In~(b), corresponding to initial state~$\ket0$,
the density distribution of the BEC is almost unchanged for the first~0.2~ms.
Then the BEC begins to change its density distribution due to shaking of the potential.
Finally, the BEC approximately reaches the target~$\ket+$ state.
For case(c),
corresponding to initial state~$\ket+$,
the density distribution of the BEC basically returns to the distribution at the initial time after~0.5 ms of evolution and in the next~0.5 ms it evolved to the~$\ket0$ state.
For case(d),
corresponding to initial state~$\ket{\text{i}}$,
we observe the same correspondence between shaking trajectory and BEC position,
but an interesting blue band,
corresponding to a narrow low-density valley,
appears in the middle of the BEC.

\begin{figure}
\includegraphics[width=0.9\columnwidth]{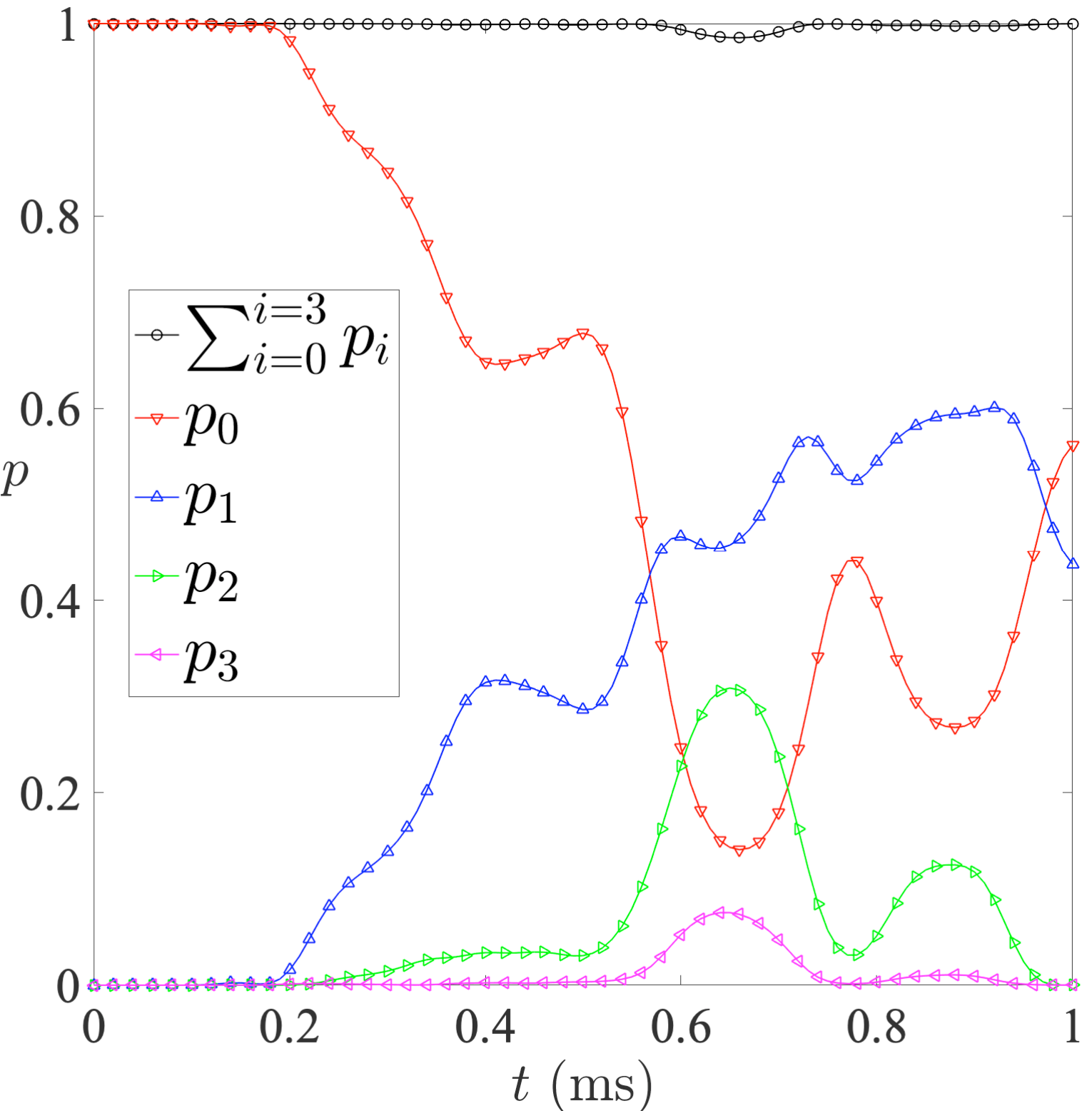}
\caption{%
    Probabilities~$\{p_i\}$ vs time~$t$
    and their time-dependent sum
    as described in the legend.%
}
\label{fig:populationdistribution}
\end{figure}

Avoiding the transition of atoms to higher energy levels is important for a feasible physical realization
because transitioning via higher levels is too difficult to control in practice.
In Fig.~\ref{fig:populationdistribution}
we plot probabilities of the first four energy levels and the sum for initial state~$\ket0$.
We can see that the sum is almost~1 and the simulation results tell us that the minimum sum is~0.985, which occurs at $t=0.66$~ms.
Here $p_0$ starts at~1 as the initial state is~$\ket0$ and remains unchanged until time~$t\approx0.2$~ms;
then this probability exhibits two regions of declining.
Between these two regions a short weak rise happens.
Finally, the probability climbs over a small hill to reach final result of $p_0=0.56$.

The behavior of~$p_1$ is just the opposite of~$p_0$, which is easy to understand because their sum needs to be close to~1.
The plot for~$p_2$ shows us that the third level plays an important role in this control,
and the plot for~$p_3$ shows us that the forth level is just a perturbative.
The crossing point of~$p_0$ and~$p_1$
occurs at~$t\approx0.58$~ms;
after that time levels three and four play important role.
To figure out which frequency component causes the BEC to transition to the second excited state,
we remove the 2 kHz frequency component contained in the optimal control trajectory;
we find that the probability of the BEC transition to the second excited state is greatly suppressed,
but at the same time, the gate fidelity is also greatly reduced.

\begin{figure}
\includegraphics[width=0.9\columnwidth]{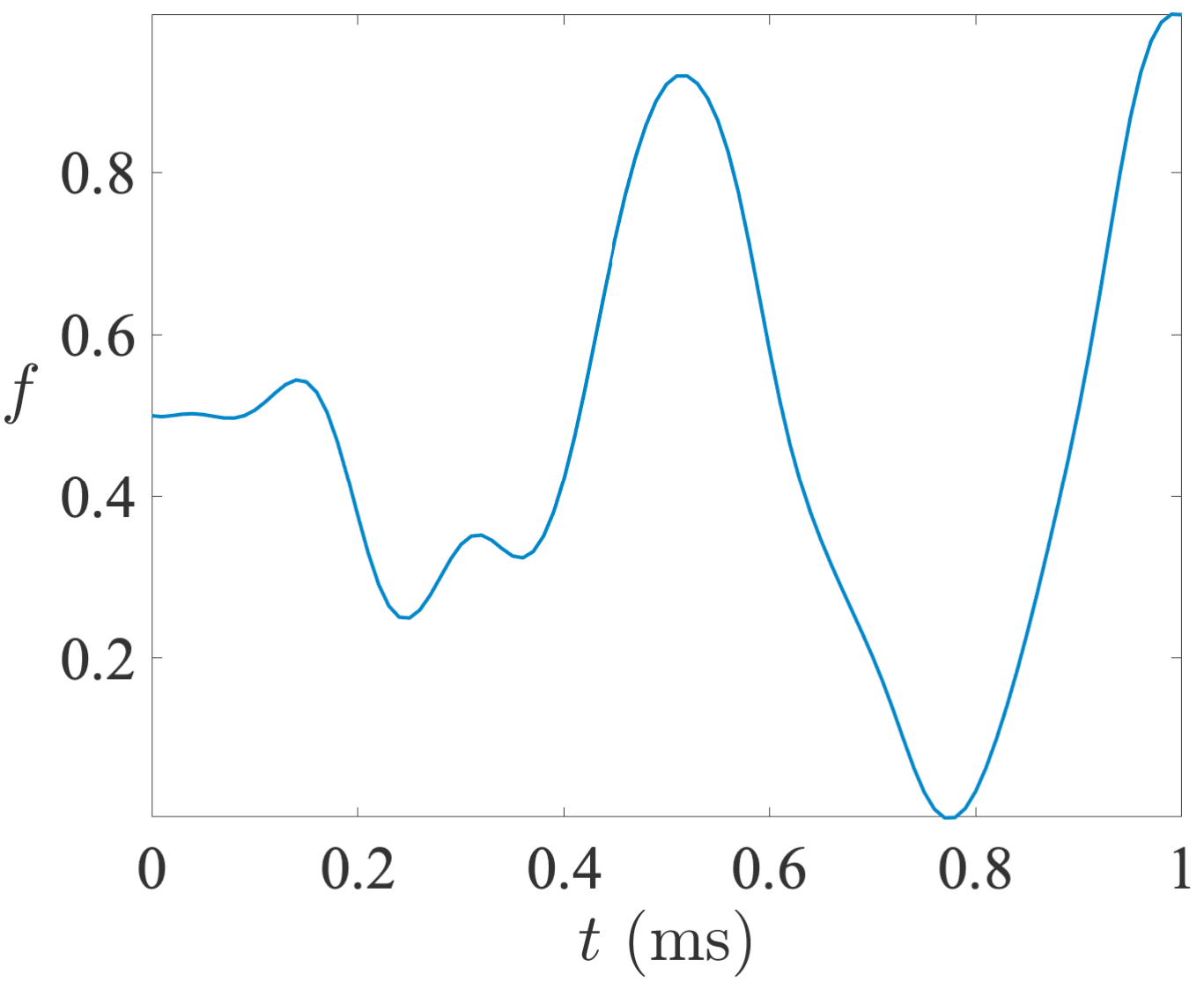}
\caption{%
    Fidelity as a function of control time.
    The BEC is initially in the~$\ket{0}$ state;
    the optimal control trajectory realizes the Hadamard gate.
    At the end of the control, the state fidelity reaches 99.45\%.
}
\label{fig:plus_fidelity}
\end{figure}

Although the~$p_0$ of the~$\ket+$ state is 0.5, and the~$p_0$ of the final state here is 0.56, the fidelity of the final state and the~$\ket+$ state here reaches 99.45\%.
In Fig.~\ref{fig:plus_fidelity}, we plot the fidelity as a function of control time.
In Table~\ref{tab:wavefunction} we also show the complex components of the final wave function of the BEC on the first four bases after applying the optimal control trajectory.
\begin{table}[]
\begin{tabular}{|c|c|c|c|c|}
\hline
 i & $\text{Re}(\braket{i|\psi})$ & $\text{Im}(\braket{i|\psi})$ \\
 \hline
 0 & 0.7274 & 0.1809\\
 \hline
 1 & 0.6498 & 0.1231\\
 \hline
 2 & -0.0223 & 0.0022\\
 \hline
 3 & -0.0090 & 0.0018\\
 \hline
\end{tabular}
\caption{Complex coefficients of the first four energy levels of the final wave function when the BEC is initially in the~$\ket{0}$ state and controlled by Hadamard gate trajectory.}
\label{tab:wavefunction}
\end{table}

\subsection{Ramsey interferometry}

We now show results of the Ramsey interferometer simulated for the case that the Hadamard gate in~\S\ref{subsec:Htrajectories} is used.
The Ramsey interferometer we simulate here consists of two Hadamard gate with free evolution in between~(for detailed information about Ramsey interferometry,
please refer to Appendix~\ref{appe:ramseyinterferometry}).
We show the trajectory of the state on the Bloch sphere,
first with~$\ket0$ and then after the first nonlinear Hadamard gate,
and then we show how the state evolves during free nonlinear evolution,
and, finally, after the last nonlinear Hadamard gate.
Finally,
we show the resultant contrast of this nonlinear Ramsey interferometer.

\begin{figure}
\includegraphics[width=0.9\columnwidth]{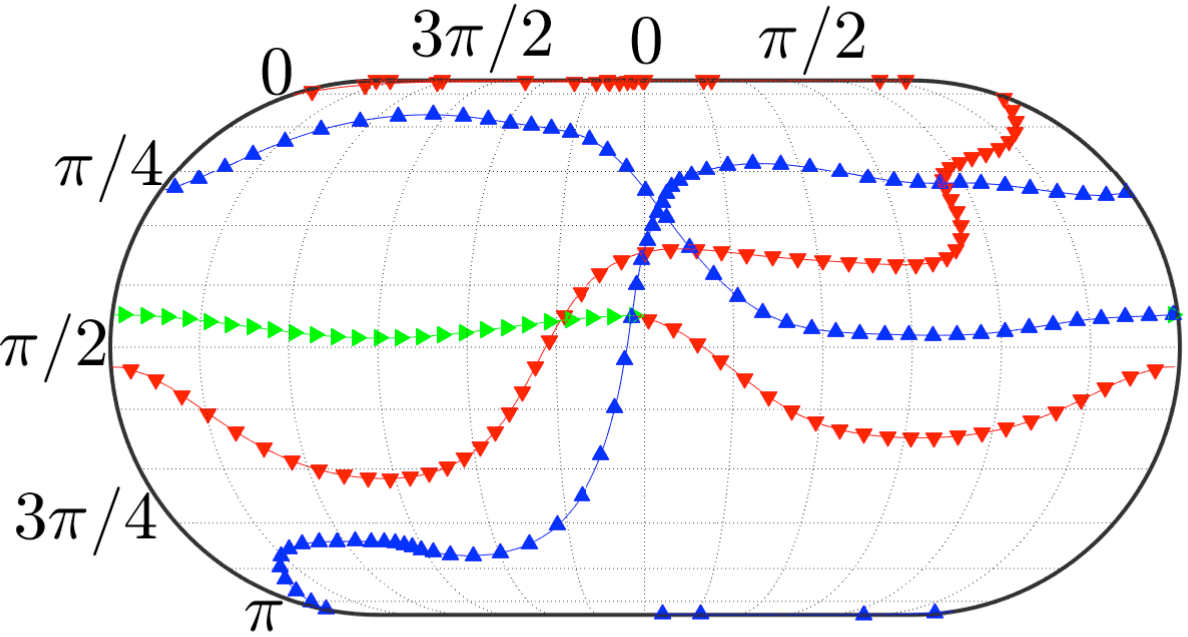}
\caption{%
    Visualization of BEC evolution on~$\mathcal{S}^2$ when the free-evolution time is half a cycle of the interferometer for the nonlinear coefficient~$g=h\times 223$~Hz~$\mu$m.
    The red downward, green rightward, and blue upward triangles represent the evolution process for the cases
     (i)~$\ket0\to\ket+$, (ii)~$\ket+\to\ket-$, and (iii)~$\ket-\to\ket1$, respectively.
     During evolution,
     transitions to higher energy levels,
     as shown in Fig.~\ref{fig:populationdistribution}, quantum states,
     are not always restricted to~$\mathcal{S}^2$.
}
\label{fig:ramseybloch}
\end{figure}

In Fig.~\ref{fig:ramseybloch}, we plot evolution of the BEC as a trajectory on a projection of the Bloch sphere.
Initially, the BEC is in the ground state~$\ket0$,
and the nonlinear Hadamard gate transfers the BEC from~$\ket0$ to an approximation of the superposition~$\ket+$,
as shown in red.
During the control process,
the initial state~$\ket0$ leaves the north pole,
which represents this initial state,
and moves ``south'' where we see that the trajectory remains in the far ``north'' and wraps around~$\mathcal{S}^2$.
After hovering near the north pole,
the trajectory first moves south and then
moves left (``west'') 
until it reaches the left side of our projection and reappears on the right side of the projected~$\mathcal{S}^2$.
Finally, the trajectory shows that the state arrives quite near to the target:
~$\ket+$ state.

After the BEC reaches a state that is close to~$\ket+$,
free evolution takes place until the second nonlinear Hadamard gate is applied.
Free evolution for the BEC system over a fixed time is shown by the green line in Fig.~\ref{fig:ramseybloch}.
Intuitively, for linear quantum mechanics,
the BEC-state trajectory would follow the ``equator'',
but Fig.~\ref{fig:ramseybloch} shows otherwise.
Similarly, if the nonlinear coefficient is~0,
the state would evolve strictly along the equator (not shown).
Here, we fix the free-evolution time so that the BEC state representation on~$\mathcal{S}^2$ moves from the left boundary (``far west'')
to the sufficiently near the right boundary (``far east''),
which, due to the nature of the projection,
are the same line (of ``longitude'').
That is, free evolution leads to the initial state accumulating only (approximately) a relative phase between~$\ket0$ and~$\ket1$ in the superposition.

Next, after free evolution,
the second, and last,
nonlinear Hadamard gate is applied followed by state readout.
This final evolution due to the nonlinear Hadamard gate is shown as the blue line in Fig.~\ref{fig:ramseybloch}.
We see that the trajectory representing the state first evolves from right to left (to the west) in the northern hemisphere and,
ultimately, arrives near~$\ket1$ in the southern hemisphere as desired for this choice of free-evolution time.
The readout is simply a projection of the resultant state onto the sub-basis $\{\ket0,\ket1\}$.

\begin{figure}
\includegraphics[width=0.9\columnwidth]{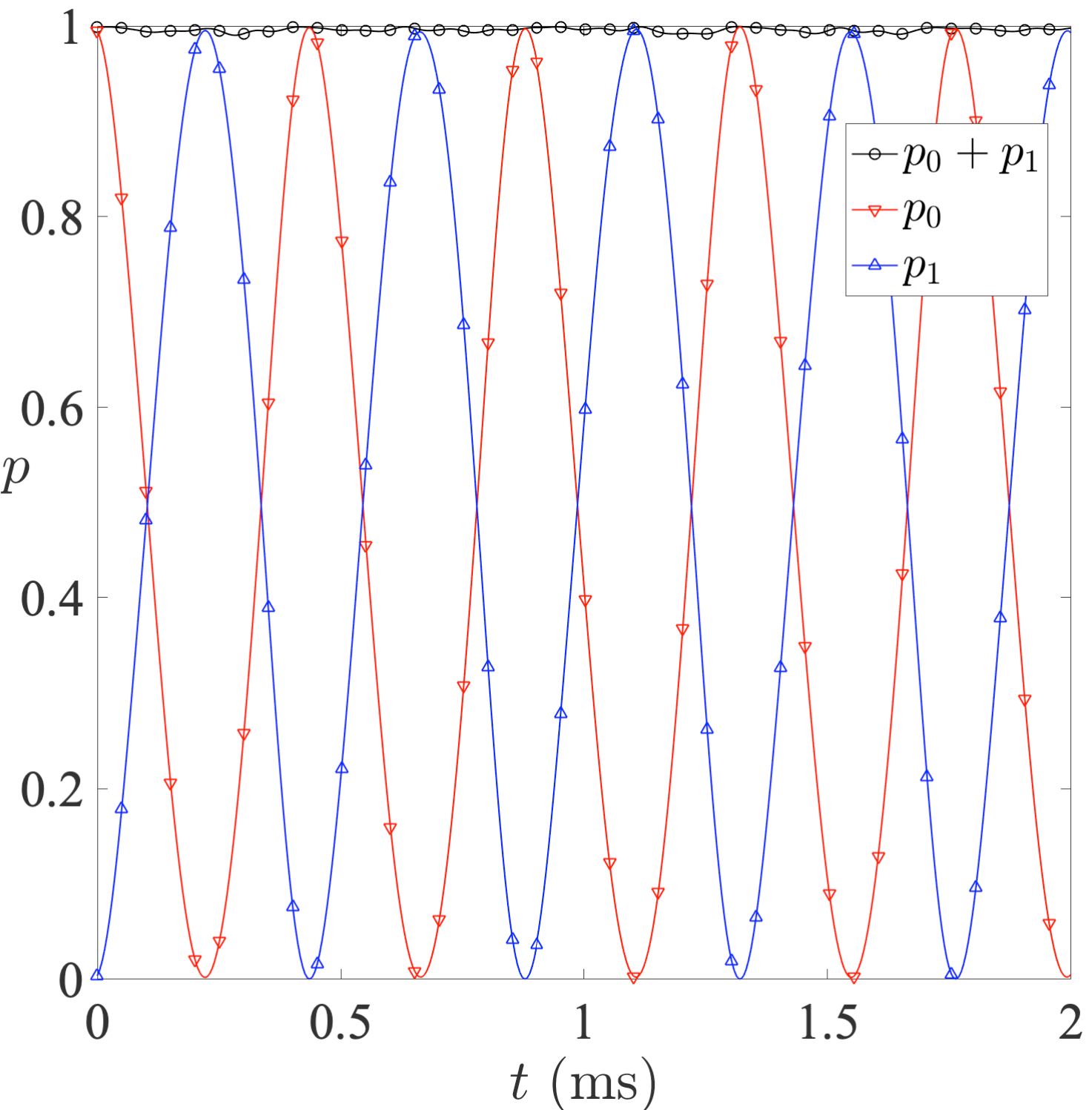}
\caption{%
    Probabilities~$\{p_{0,1}\}$ vs time~$t$
    and their time-dependent sum
    as described in the legend.%
}
\label{fig:contrast}
\end{figure}

We consider different free-evolution times and show the resultant contrast for the continuum of these evolution times.
Figure~\ref{fig:contrast}
is a plot of $i^\text{th}$-level atomic-state probability vs time~$t$
and shows different fixed time and the corresponding population distribution of atoms in the final read out stage.
The contrast is
\begin{equation}
\label{eq:contrast}
\mathcal{C}(p_i)
:=\frac{\text{max}(p_i)-\text{min}(p_i)}{\text{max}(p_i)+\text{min}(p_i)},
\end{equation}
for~$p_i$ the probability for being measured in $\ket{i}$~(\ref{eq:p_i}).
We observe in Fig.~\ref{fig:contrast}
that the maximum achievable contrast is
\begin{equation}
    \mathcal{C}(p_0)\approx99.63\%,\, \mathcal{C}(p_1)\approx99.96\%.
\end{equation}
This contrast far exceeds the earlier maximum of 97\% contrast for a nonlinear Ramsey interferometer~\cite{van2014}.
We see from Fig.~\ref{fig:contrast}
that the frequency gap between~$\ket0$ and~$\ket1$ is about 2.27~kHz,
which is quite close to the frequency gap obtained by solving the GPE, which yields a frequency gap of 2.26~kHz.
These frequency gaps are slightly smaller than the single-particle frequency splitting 2.28~kHz for the nonlinear coefficient being zero,
i.e., the linear quantum mechanical case.
Our control trajectory shows excellent results as the BEC has only a 0.97\% probability of evolve beyond~$\ket1$
during the control process.
\section{Discussion }
\label{sec:discussion}

In this section,
we present an overview of our results
and interpretation.
We first discuss the importance of choosing a feasible potential,
which we treat as quartic and then search for appropriate parameters,
for realizing nonlinear quantum gates.
As this technique involves searching the frequency domain,
we discuss how we restrict frequency bandwidth by truncation techniques.
Finally, we discuss how to use our gate design to realize a nonlinear version of a Ramsey interferometer,
which is implemented using our nonlinear Hadamard gate twice with nonlinear free evolution in between.

We have seen that computing feasible nonlinear quantum gates is difficult,
not just due to high computational demands,
but also because even the notion of a quantum gate must be generalized to accommodate nonlinearity for which the superposition principle fails.
We have addressed this conceptual challenge
and then developed mathematical,
and concomitant numerical,
methods to obtain satisfactory parameters for a quartic potential.
We then assess each candidate potential
by evaluating its average gate fidelity,
appropriately defined for nonlinear quantum gates;
this average gate fidelity informs us as to the feasibility of the candidate potential
as well as enabling comparison of relative efficacy for the candidate potential.
If a candidate shaking trajectory is infeasible,
we continue our mesh search for feasible potential parameters.
After determining appropriate potential parameters,
we then use a standard global search algorithm to find a feasible control trajectory to implement the target nonlinear quantum gate.

In order to perform an efficient and effective search, we implement a standard global search algorithm in the frequency domain subject to a carefully determined bandwidth cutoff,
based on computing the spectral averaging function.
Plotting this function then shows us approximately how much bandwidth restriction is acceptable insofar as not eliminating important features of this function by an overzealous truncation.
Optimizing this bandwidth cutoff is somewhat subjective but is tested by trying different cutoffs to assess convergence for larger bandwidth choices.
Here we tested two cases:
the three-frequency cutoff and the nine-frequency cutoff.
For our purposes we observe that average gate fidelity for the nine-frequency case is better than the three-frequency case so truncating for three frequencies would be ill advised in this case.

We have used global search to compute a feasible shaking trajectory for implementing a nonlinear Hadamard gate,
and then we use this resultant nonlinear gate as part of our simulation of nonlinear Ramsey interferometer.
Our nonlinear Ramsey interferometer comprises two nonlinear Hadamard gates with free evolution occurring in between.
To analyze nonlinear Ramsey interferometry,
we represent the evolving BEC state on the Bloch sphere,
restricted to the ground and first excited state levels,
over the whole process from when the BEC starts in the ground state
until the end.
In addition,
we have calculated contrast of the interferometer,
and our results show high contrast,
which further confirms that our nonlinear Hadamard gate is quite effective.

Here we have considered radial degrees of freedom in the splitting direction. 
We did not consider the longitudinal degrees of freedom,
nor the other transverse direction,
which have significantly different energy levels and thus will not contribute. 
These aspects have been extensively discussed in the original experiments studying similar excitations~\cite{bucker2011,bucker2013,van2014,borselli2021two}.

\section{Conclusions}
\label{sec:conc}
In conclusion,
we have introduced a rigorous foundation for treating nonlinear quantum gates,
for both interferometry and quantum information processing,
and applied our framework to the case of Bose-Einstein condensates,
whose dynamics is governed by a nonlinear Schr\"{o}dinger equation.
We apply this framework to determining feasible parameters of both the quartic potential that confines the Bose-Einstein condensate and how this potential is shaken.
Using global optimization methods,
we are able to determine a shaking trajectory such that an average gate fidelity of 99.21\% can be achieved for a nonlinear Hadamard gate, and using this shaking trajectory twice,
with nonlinear free evolution between the two gates,
the effective contrast for a nonlinear version of Ramsey interferometry reaches 0.99.
We have thus established a method for devising, assessing, and applying nonlinear gates to a trapped Bose-Einstein condensate,
which could be tested experimentally.

Our work includes a definition of nonlinear quantum gates,
which combines the salient features of linear quantum gates but connects with nonlinear quantum mechanics, for which nonlinear unitary evolution only preserves the norm but not the inner product.
Our control method requires sampling over a wide range of initial states;
this added complexity is due to the nonlinearity,
which violates superposition rules of linear quantum mechanics.
Our approach to optimizing trapping-potential parameters and computing a feasible shaking trajectory is achieved by searching for time-dependent trajectories but restricted by a bandwidth cutoff based on first analyzing frequency-domain support for the spectral averaging function.
Then, subject to this restriction,
global search methods are employed to devise feasible nonlinear quantum gates.
In particular we devise a fast high-fidelity nonlinear Hadamard gate,
and use this gate to demonstrate high-contrast nonlinear Ramsey interferometer.
multi body quantum systems.
Other types of quantum gates can also be implemented by our strategy.

Our work sets the stage for some interesting future investigation.
Given our definition of a nonlinear quantum gate,
we could ask about nonlinear quantum computing and how, in particular,
a computer founded on interacting trapped Bose-Einstein condensates could operate
and whether nonlinearity would be advantageous,
disadvantageous, or neither.
Here we have studied a single-qubit gate,
but quantum control for a two-qubit gate is important for quantum information processing;
we expect similar quantum-control techniques as developed here would be applied in that more complicated case,
but tractability of the computational control problem is currently an open question.
Another direction is to implement this scheme in the laboratory,
which would likely require further tuning of our model to accommodate nonideal features such as non quartic features of the potential and ramping and latency issues in the time-dependent shaking control operation.

\begin{acknowledgments}
We acknowledge China’s 1000 Talent Plan and NSFC (Grant No. 11675164) and Anhui Initiative in Quantum Information Technologies for support.
The simulations were carried out at the Supercomputing Center of the University of Science and Technology of China.
The work at TU-Wien was supported by the EU’s Horizon 2020 program under the M. Curie Grant No. 765267 (QuSCo), and by the Wiener Wissenschafts- und TechnologieFonds (WWTF), project No. MA16-066 (SEQUEX).
\end{acknowledgments}

\appendix
\section{BEC stationary states by minimization}
\label{appe:bec}
We explain stationary states of the BEC by casting this as a minimization problem.
First, we discuss the equation that the BEC stationary states need to satisfy.
The solution of the BEC stationary states involves nonlinear eigenvalue problems.
Finally, we define ground and excited BEC states.

We discuss stationary states of the BEC.
In order to find the stationary state~(\ref{eq:1dham}),
we write~\cite{bao2006}
\begin{equation}
\label{eq:realtime}
\psi(x;t)
    =\text{e}^{-\text{i}\mu[\phi]t}\phi(x),
\end{equation}
where~$\mu[\phi]$ is the chemical potential functional
of the BEC wave function and~$\phi(x)$ a normalized Lipschitz-continuous function.
We denote the set of all such~$\phi(x)$ by~$\Phi$.
By inserting Eq.~(\ref{eq:realtime}) into Eq.~(\ref{eq:1dham}),
we obtain
\begin{equation}
\label{eq:stationary}
\mu[\phi]\phi
=-\frac{\hbar^2}{2m}\partial_{xx}\phi+V(x)\phi+g|\phi|^2\phi,
\end{equation}
which BEC stationary states must satisfy.
From Eqs.~(\ref{eq:energy}) and~(\ref{eq:realtime}),
we see that the energy of the BEC for the stationary state~$\phi$ simplifies to 
\begin{equation}
\label{eq:energy'}
E[\phi]
=\int_\mathbb{R}\text{d}x
\left[\frac{\hbar^2}{2m}
\partial^2_x\phi
+V(x)\phi^2
+\frac{g}2\phi^4\right],
\end{equation}
which is time-independent as we are focused here on stationary states.
We now have explained the standard approach to stationary states and energies,
but superpositions and unitary operators are problematic due to nonlinear evolution,
so we now elaborate on the mathematical context for  nonlinear evolution with such states and energies.

We discuss essential properties regarding the nonlinear eigenproblem.
Stationary states of the BEC pertain to a nonlinear equation as standard linear-operator methods are inapplicable~\cite{ruhe1973,guillaume1999}.
The stationary state of the BEC is actually a nonlinear eigenstate~(\ref{eq:stationary}),
and the corresponding eigenvalue is
\begin{align}
\mu[\phi]
=&\int_\mathbb{R}\text{d}x\left[\frac{\hbar^2}{2m}\left|\partial_x\phi\right|^2+V(x)\left|\phi\right|^2+g|\phi|^4\right]\nonumber\\
=&E(\phi)
+\int_\mathbb{R}\text{d}x\frac{g}{2}\left|\phi\right|^4,
\end{align}
which is the chemical potential of the BEC in the state~$\phi$.

We discuss the definition of the ground state and the first excited state of the BEC.
The ground state~$\phi_0$ of the BEC satisfies the minimization condition~\cite{bao2006}
\begin{equation}
\label{eq:E0}
E_0 := E[\phi_0]
=\min_{\phi\in\Phi} E[\phi],
\end{equation}
and the first excited state~$\phi_1$ of the BEC satisfies the minimization condition
\begin{equation}
\label{eq:E1}
E_1 := E[\phi_1]=\min_{\phi\in\Phi\setminus\{\phi_0\}} E[\phi].
\end{equation}
Other states satisfying Eq.~(\ref{eq:stationary}), but whose energies~(\ref{eq:energy'}) exceed~$E_0$,
are called excited states.

\section{Optimal control of BEC shaking}
\label{appe:oct}

We review relevant work on optimal control of a BEC,
and,
in this appendix,
our main reference is Hohenester's 2014 study~\cite{hohenester2014}.
We discuss the shaking model for an effective one-dimensional BEC.
Optimal control theory based on functional and Lagrangian function as the main method is also discussed.
Finally, we discuss pertinent tools in the MATLAB toolbox for optimally controlling a BEC.

We review the shaking model for an effective one-dimensional BEC.
The system under consideration is a BEC in a trap, where the dynamics takes place in one dimension.
The control part is achieved by shaking the trap.
The physical realization of this model can be implemented on an atomic chip~\cite{trinker2008};
shaking can be achieved by changing the electrical current in the wires.
 
We now describe Hohenester's optimal control method, which is based on treating the Lagrangian function and employing variational calculus.
Given the initial state and the target state of the BEC as well as the shape of the potential and the control time~$T$,
the task is to design a time-dependent shaking trajectory that transforms the initial state to the target state.
Due to the constraints,
a optimal control sequence is not guaranteed to exist,
but Hohenester's optimal control method attempts to determine the best sequence for shaking the BEC.
Hohenester's method proceeds according to the following steps.
First, a guessed shaking trajectory is applied,
then, the cost function related to the shaking trajectory is constructed,
finally, the Lagrange function and functional derivatives are used to determine the optimal way to change the guessed shaking trajectory.
Details of the Lagrangian functional based methods are well studied~\cite{hohenester2014,hohenester2007optimal}.

We now describe Hohenester's toolbox used to design control sequences, OCTBEC~\cite{hohenester2014}.
OCTBEC is a MATLAB toolbox that provides a MATLAB class known as \textsc{optimize},
where we use \textsc{camelCase} to denote computer functions,
in order to perform calculation of optimal control sequences.
For a single input state and target state,
the toolbox yields an optimal shaking trajectory,
which is treated as the best as the optimum.
This toolbox employs a gradient-type algorithm to determine the search direction of the control sequences.
Two gradient-type algorithms are implemented in this toolbox,
which are the nonlinear conjugate gradient and quasi-Newton optimization~\cite{hohenester2014}.
The Crank–Nicolson method~\cite{crank1996},
used for numerical simulation of BEC evolution,
is incorporated into this toolbox through a MATLAB function,
known as~\textsc{solve}.

\section{Quantum information processing with BECs}
\label{appe:qipbec}

We review different schemes in quantum computing with BECs.
The main difference between these schemes lies in the encoding method of the qubit and the realization of the nonlinear quantum gate.
In this subsubsection,
our main references are Shi's 2001 study~\cite{shi2001},
Hecht's 2004 study~\cite{hecht2004} and
Byrnes et al.'s 2012 study~\cite{byrnes2012}.

We now describe Shi's 2001 study~\cite{shi2001}.
The BEC localized in the symmetric double-well potential is used as a qubit.
The pure state~$\ket0$ and~$\ket1$ correspond to condensate wave functions that are highly localized in the left and right well, respectively.
The condensate wave function is described by a superposition of these two states.
Single-qubit nonlinear quantum gates could be realized through Josephson-like tunneling,
and two-qubit nonlinear quantum gates could be constructed by putting together two double wells~\cite{shi2001}.
Shi's study ignores self-trapping, which was understood later by Oberthaler et al.~\cite{albiez2005direct}.
Self-trapping mitigates against this kind of single-qubit gate.

In Hecht's thesis~\cite{hecht2004},
the BEC bound in a quasi harmonic potential with atoms with two internal levels is considered.
The pure state~$\ket0$ and~$\ket1$ correspond two degenerate ground states and under certain conditions those ground states is separated from the excited levels by an energy gap.
Single-qubit gates can be performed by exploiting the Zeeman effect,
and two-qubit quantum gates can be performed by enabling tunneling between two neighboring qubit systems.

Byrnes et al.'s 2012 study~\cite{byrnes2012} consider two-level BEC system,
``such as two hyperfine levels in an atomic BEC or spin polarization states of exciton polaritons.''
The pure state~$\ket0$ and~$\ket1$ are represented as those two level.
Single-qubit gates can be performed by Hamiltonian with Schwinger boson operators,
and two-qubit quantum gates can be performed by Hamiltonian which has the form of the product of the Schwinger boson operators.

\section{Nonlinear Ramsey interferometry with BECs}
\label{appe:ramseyinterferometry}

Ramsey interferometry
is also known as the separated oscillating fields method
and is used to measure the transition frequency of a two-level atom~\cite{ramsey1950molecular}.
We first review the concept of Ramsey interferometry on both linear and nonlinear systems.
Then we review the theory of interferometers on linear systems.
Finally, we review the implementation of Ramsey interferometry on linear and nonlinear systems.
\subsection{Concept}

A Ramsey interferometer is an apparatus that employs two laser pulses,
separated in time by a fixed duration,
to an atomic gas with the goal of measuring the energy difference between the two levels.
A Ramsey interferometer has many applications,
such as for atomic clocks~\cite{santarelli1999} and for quantum simulation~\cite{cetina2016}.
Here we discuss both the case of an atomic gas and the case of a BEC.
In its simplest form, a Ramsey interferometer  comprises four stages~\cite{ramsey1956molecular}:
a~$\nicefrac\pi2$ pulse,
followed by free evolution that results in precession leading to phase accumulation,
and then another~$\nicefrac\pi2$ pulse followed finally by measurement.
Here we review these four parts in detail.

We now explain the first~$\nicefrac\pi2$ pulse for both the linear case corresponding to an atomic gas and the nonlinear case corresponding to a BEC.
In the linear case,
consider a two-level atomic system,
whose two energy levels correspond to the ground state~$\ket0$ and to the first excited state~$\ket1$ of an atom.
A~$\nicefrac\pi2$ pulse transfers population from the ground state to an equally weighted superposition state~\cite{steck2007quantum};
for linear quantum mechanics,
two concatenated, i.e., sequential,
$\nicefrac\pi2$ pulses yields a~$\pi$ pulse, which would map~$\ket0\mapsto\ket1$.

In the nonlinear case,
the~$\nicefrac\pi2$ pulse excites the BEC from the ground state to an equally weighted superposition state where the ground state is the lowest energy state of the nonlinear Hamiltonian and the first transverse-motional excited state is the discrete excitation of that energy~\cite{van2014}.
Graphically we can visualize the evolution of the state by picturing the ground states as the south pole of~$\mathcal{S}^2$, the excited state as the north pole of~$\mathcal{S}^2$ and the~$\nicefrac\pi2$ pulse as being a rotation that rotates the state at the south pole to the equator.
In the nonlinear case,
concatenating two~$\nicefrac\pi2$
pulses does not generally yield a~$\pi$ pulse, with deleterious implications for nonlinear interferometry.

Subsequent to the~$\nicefrac\pi2$ pulse,
the system then evolves freely for a fixed time.
Due to the energy difference between the ground state and the excited state,
two different atomic energy levels accumulate different phases;
the relative phase of the superposition state changes.
In the linear case,
precession corresponds to a rotation about the polar axis of~$\mathcal{S}^2$
with a constant precession rate.
In the nonlinear case the precession rate will be state-dependent and will not be constant~\cite{van2014}.

In the linear case,
the second~$\nicefrac\pi2$ pulse can be the same as the first one with the concatenation of these two pulses yielding a~$\pi$ pulse, but not so for the nonlinear case;
therefore, the second~$\nicefrac\pi2$ pulse needs to be considered carefully.
For constructing the second~$\nicefrac\pi2$ pulse,
we are motivated by interferometric considerations to regard as paramount that,
for free-evolution time being zero,
concatenating two~$\nicefrac\pi2$ pulses must excite $\ket0\mapsto\ket1$.

Whereas concatenating two~$\nicefrac\pi2$ pulses yields a~$\pi$ pulse,
concatenating the nonlinear Hadamard gate
\begin{equation}
\label{eq:Hgate}
H:\ket0\mapsto\ket+,\,
H:\ket1\mapsto\ket-,
\end{equation}
twice yields an identity operation:
$H^2=\mathds1$.
This subtlety is not important in linear quantum mechanics as long as the nonlinear Hadamard gates arise in pairs so usually~$\nicefrac\pi2$ and~$H$ gates are treated as equivalent, but the inverse property $H=H^{-1}$ is important for our approach.
Thus, for the nonlinear case,
a Hadamard pulse is used rather than a~$\nicefrac\pi2$ pulse~\cite{van2014}.
Consequently,
instead of $\ket0\mapsto\ket1$,
our concatenating two sequential nonlinear versions of~$H$
maps~$\ket0\mapsto\ket0$.
If the two sequential nonlinear~$H$ gates 
are punctuated by intervening free evolution,
the dynamics becomes complicated because the resultant qubit map is state-dependent due to nonlinearity;
experimentally, this complication is circumvented by restricting states to polar states~$\ket0$ and~$\ket1$ or to the equator states $\ket0+\text{e}^{\text{i}\varphi}\ket1$
using the~$\mathcal{S}^2$ representation~\cite{van2014}.

The last step of the Ramsey interferometer is measurement.
In the two-level atomic system,
one method of measurement is
when the atomic beam completes the interference process and enters the measurement cavity;
the number of atoms in the ground state and excited state can be counted by counting ions which are generated by the ionization of atoms on the hot tungsten wire~\cite{cronin2009optics}.
In the BEC case,
populations of the ground state and first excited state of the BEC are inferred from the evolution of the momentum density, which is obtained by time-of-flight images~\cite{van2014}.

\subsection{Theory}

As background to our analysis of nonlinear Ramsey 
interferometry,
we first elaborate here how this system works in the case of linear quantum mechanics.
We explain based on a model for a single atom interacting twice with a single-mode pulse~\cite{scully1999quantum}.
We first discuss the~$\nicefrac\pi2$ pulses,
then the dynamics,
and, finally probabilities for measuring the atom at the two distinct energy levels defining the qubit.

Now we explain how to implement a~$\nicefrac\pi2$ pulse in the actual physical system for linear quantum mechanics.
When the single-mode pulse is at resonance with the atom,
the resultant unitary evolution is
\begin{equation}
    \cos\left(\nicefrac{\omega t}2\right)\sigma_0-\text{i}\sin\left(\nicefrac{\omega t}2\right)\sigma_x,\,
    \sigma_i\in\mathcal{M}_2(\mathbb{C})
\end{equation}
with~$\mathcal{M}_2(\mathbb{C})$ shorthand for~$2\times2$ complex matrices.
Here,
$\sigma_x$ is the Pauli~$X$ matrix and~$\Omega(t)$ is the time-dependent Rabi frequency~\cite{steck2007quantum}
with time dependence suppressed for convenience.
If the atom is initially in the ground state~$\ket0$,
when the interaction time of the single-mode field and the atom satisfies~$\int\text{d}t\Omega(t)=\pi/2$,
the atom evolves to a superposition state and this pulse is called a~$\nicefrac\pi2$ pulse.

The~$\nicefrac\pi2$ pulse is the key experimental challenge for Ramsey interferometry, 
and now we discuss free evolution between these two pulses.
Initially, the atom is prepared in the ground state~$\ket0$.
The first~$\nicefrac\pi2$ pulse is applied to map
$\ket0\mapsto\ket0-\text{i}\ket1$,
and then free evolution occurs for time~$t$ as a precession phenomenon;
$t$ might or might not be controllable depending on context.
For frequency difference~$\omega$
between the atomic ground and excited states,
the electron state after time~$t$ is~$\ket0-\text{ie}^{-\text{i}\omega t}\ket1$.

The final~$\nicefrac\pi2$ pulse ends the quantum process prior to the ultimate measurement.
After removing an unobservable global phase,
the final electronic state is~$\sin(\nicefrac{\omega t}2)\ket0-\cos(\nicefrac{\omega t}2)\ket1$.
As a special case,
a trivial free-evolution time of $t=0$, the atom evolves according to $\ket0\mapsto\ket1$.
By changing the precession time, different final states can be obtained,
with the precession time leaving its signature on the final state.

The final step of Ramsey interferometry is to measure whether the atom is excited or not.
In practice, by repeating over many atoms,
we obtain the population difference between ground and excited states.
Thus, mathematically, we measure in the~$\{\ket0,\ket1\}$ basis with the result that the probability for the atom being in~$\ket0$ is~$\sin^2(\nicefrac{\omega t}2)$ and in~$\ket1$ is~$\cos^2(\nicefrac{\omega t}2)$

\subsection{Experiments}

We now summarize experiments concerning the implementation of Ramsey interferometry for both linear and BEC systems.
We discuss the realization of~$\nicefrac\pi2$ pulses.
Then we discuss free evolution and finally how measurement is performed in different experiments.

We review the standard experimental realization of Ramsey interferometry~\cite{wu2013accurate}.
In Ramsey interferometry using $^{87}$Rb atoms,
the measurement yields information about the frequency difference between the F=1 and F=2 hyperfine levels.
The~$\nicefrac\pi2$ pulse is achieved either by a direct microwave transition between the two levels or indirectly by a Raman transition between the two levels.
During the free evolution,
the field is turned off.
The resultant energy level is measured by absorption imaging.

We review the experimental realization of Ramsey interferometer on BEC system~\cite{van2014}.
Instead of two hyperfine levels,
the two-level system is achieved by accessing only the two lowest motional states in the potential that confines the BEC.
The~$\nicefrac\pi2$ pulse is achieved by shaking the potential,
which in turn is achieved by varying the magnetic field.
For the nonlinear Ramsey interferometer,
we use an alkali atom,
for example rubidium,
and an alkali atom has a magnetic moment due to the spin and orbital motion of of electrons outside the nucleus;
therefore, the atoms can be controlled by varying the magnetic field~\cite{reichel2011atom}.

Free evolution is achieved by ceasing the shaking for a fixed duration.
After free evolution,
the potential is shaken again to achieve another~$\nicefrac\pi2$ pulse;
due to nonlinearity, the shaking trajectory can differ for the second
$\nicefrac\pi2$
pulse compared to the first.
At the measurement stage,
populations of the ground state and first excited state of the BEC are inferred from evolution of the momentum density, which is obtained by time-of-flight images.

\bibliography{biblio}
\end{document}